\documentclass[%
 reprint,
 amsmath,amssymb,
 aps,pra,
]{revtex4-2}

\usepackage{color}
\usepackage{graphicx,float}
\usepackage{dcolumn}
\usepackage{bm}
\usepackage{xcolor}
\usepackage{soul}
\usepackage{upgreek}

\usepackage{balance}

\usepackage{epstopdf}
\usepackage[caption=false, justification=centering, singlelinecheck=false]{subfig}
\usepackage{xcolor} 
\usepackage{soul,color}
\usepackage[capitalise,nameinlink]{cleveref}
\usepackage{placeins}

\begin{document}

\preprint{APS/123-QED}

\title{Effects of polydispersity and concentration on elastocapillary thinning of dilute polymer solutions }

\author{Vincenzo Calabrese}
\author{Amy Q. Shen}
\author{Simon J. Haward}
\affiliation{Okinawa Institute of Science and Technology Graduate University, Onna-son, Okinawa, 904-0495, Japan
}%

\date{\today}

\begin{abstract}

The self-thinning of liquid bridges under the action of capillarity occurs in widespread processes like jetting, dripping, and spraying, and gives rise to a strong extensional flow capable of stretching dissolved polymers. If the resulting elastic stress exceeds the viscous stress, an exponential `elastocapillary' (EC) thinning regime arises, yielding a timescale $\tau_{EC}$ that is commonly considered equivalent to the longest relaxation time of the polymer $\lambda$. A longstanding question is why $\tau_{EC}$ depends strongly on the polymer concentration, even at high dilutions where $\lambda$ should be constant in theory. To date this is understood in terms of intermolecular interactions that arise due to `self-concentration' effects as polymers stretch. However, $\lambda$ depends on the polymer molecular weight $M$, and we show how the concentration dependence of $\tau_{EC}$ can be explained by considering the molecular weight distribution (MWD) inherent in real polymer samples, without the need to invoke self-concentration. We demonstrate this by blending low-$M$ and high-$M$ polymer samples with narrow MWDs at dilute concentrations and in different proportions, and by measuring $\tau_{EC}$ for each blend in capillary thinning experiments. Through a simple model that qualitatively reproduces the experimental results, we show how elastic stresses generated by the polymer build up prior to the EC regime due to the sequential stretching of progressively decreasing molecular weight species in the MWD. Since the elastic stress generated by each species depends on its concentration, the fraction of the MWD that is required to stretch in order to induce the EC regime depends on the total polymer concentration $c$ in the solution. For higher $c$ the EC regime is induced by stretching of a higher-$M$ (longer $\lambda$) fraction of the MWD, and results in a longer measurement of $\tau_{EC}$. Our results have significant implications for the application of capillary thinning measurements to extensional rheometry, for the interpretation of such measurements, and for the understanding of elastocapillary thinning dynamics in general.

\end{abstract}

\maketitle

\section{\label{Int}Introduction}

The self-thinning and breakup of liquid bridges under the action of capillarity is ubiquitous in widespread natural and industrial processes involving both simple Newtonian fluids (like water) and more complex non-Newtonian fluids (e.g., solutions of proteins, DNA, or synthetic polymers). Important examples include the fluid breakup and fragmentation processes in spraying and coating applications \cite{Keshavarz2016,Keshavarz2020}, dripping \cite{Amarouchene2001,Ingremeau2013}, fiber-spinning \cite{Xue2019,Xu2022}, ink-jet printing \cite{Hutchings2013,McIlroy2013,Lohse2022}, misting and aerosolization (including during sneezing and speech) \cite{Wei2015,Scharfman2016,Abkarian2020,Bourouiba2021}. The inertialess  capillary-driven thinning of Newtonian liquid bridges is controlled by a viscocapillary (VC) stress balance and is linear in time \cite{Papageorgiou1995,McKinley2000}. For polymer solutions, the squeezing of the fluid neck by the capillary pressure can induce additional elastic stresses associated with the stretching of polymers, resulting in a regime of exponential thinning controlled by an elastocapillary (EC) stress balance \cite{Entov1997,Anna2001b,Clasen2006}. Measurements of the rate of exponential thinning in the EC regime are widely used to extract rheological properties, such as characteristic `relaxation times', of polymeric fluids \cite{Anna2001b,Campo2010,Nelson2011,Bhattacharjee2011,Dinic2015,Keshavarz2015,Sharma2015,Mathues2018,Rajesh2022,Gaillard2023}. 

Experimentally it is observed that characteristic timescales $\tau_{EC}$ measured in the EC regime depend upon the concentration $c$ of polymer in solution, even well below the equilibrium polymer overlap concentration $c^*$ where interpolymer interactions should be negligible and the relaxation time $\lambda$ should be independent of $c$ \cite{McKinley2005,Tirtaatmadja2006,Clasen2006,Mathues2018,Dinic2019,Colby2023,Soetrisno2023}. In a 2005 review article, this unexpected concentration-dependence of $\tau_{EC}$ was cited as an ``outstanding remaining challenge'' to the understanding of elastocapillary thinning dynamics \cite{McKinley2005}. Since then, it has become accepted that intermolecular interactions become important as coiled polymer chains unravel and stretch in the thinning fluid neck, affecting the determination of $\tau_{EC}$ even if the fluid would normally be considered dilute at equilibrium. The increasing dimension of the coil as the polymer unravels is understood as leading to an effectively reduced overlap concentration, a so-called `self-concentration' effect. The implication is that, in order to obtain a concentration independent value for $\tau_{EC}$, the fluid must be ‘ultradilute' at equilibrium, such that intermolecular interactions remain negligible even when the polymer becomes highly stretched \cite{Harrison1998,Clasen2006,Dinic2020}.

Although intermolecular interactions will clearly influence polymeric relaxation times beyond a certain concentration (normally considered to be $c^*$), in this paper we consider the role of a polydisperse molecular weight distribution (MWD) and its possible influence on the determination of $\tau_{EC}$. Polydispersity in molecular weight results in a wide spectrum of relaxation times, significantly affecting the mechanical and rheological properties of polymers \cite{Middleman1967,Bersted1979,Gentekos2019}. Experimental studies (e.g., \cite{Plog2005,Kim2019,Soetrisno2023}) clearly show that polydispersity influences EC thinning dynamics. 
It has also been shown how the addition of a high molecular weight polymer to a lower molecular weight solution (making a bimodal MWD) can stabilize threads of elastic fluids, retarding their breakup in fiber-spinning applications \cite{Palangetic2014,Merchiers2022}. But perhaps the most influential interpretation of the role of MWD on the EC thinning of polymer solutions stems from an early theoretical analysis of a multimode finitely extensible nonlinear elastic (FENE) fluid \cite{Entov1997}. From this analysis it is widely thought that the characteristic time $\tau_{EC}$ corresponds to the relaxation time of the longest, or highest molecular weight, polymers in the sample. 


\begin{figure}[t!]
    \centering
    \includegraphics[width=8.0cm]{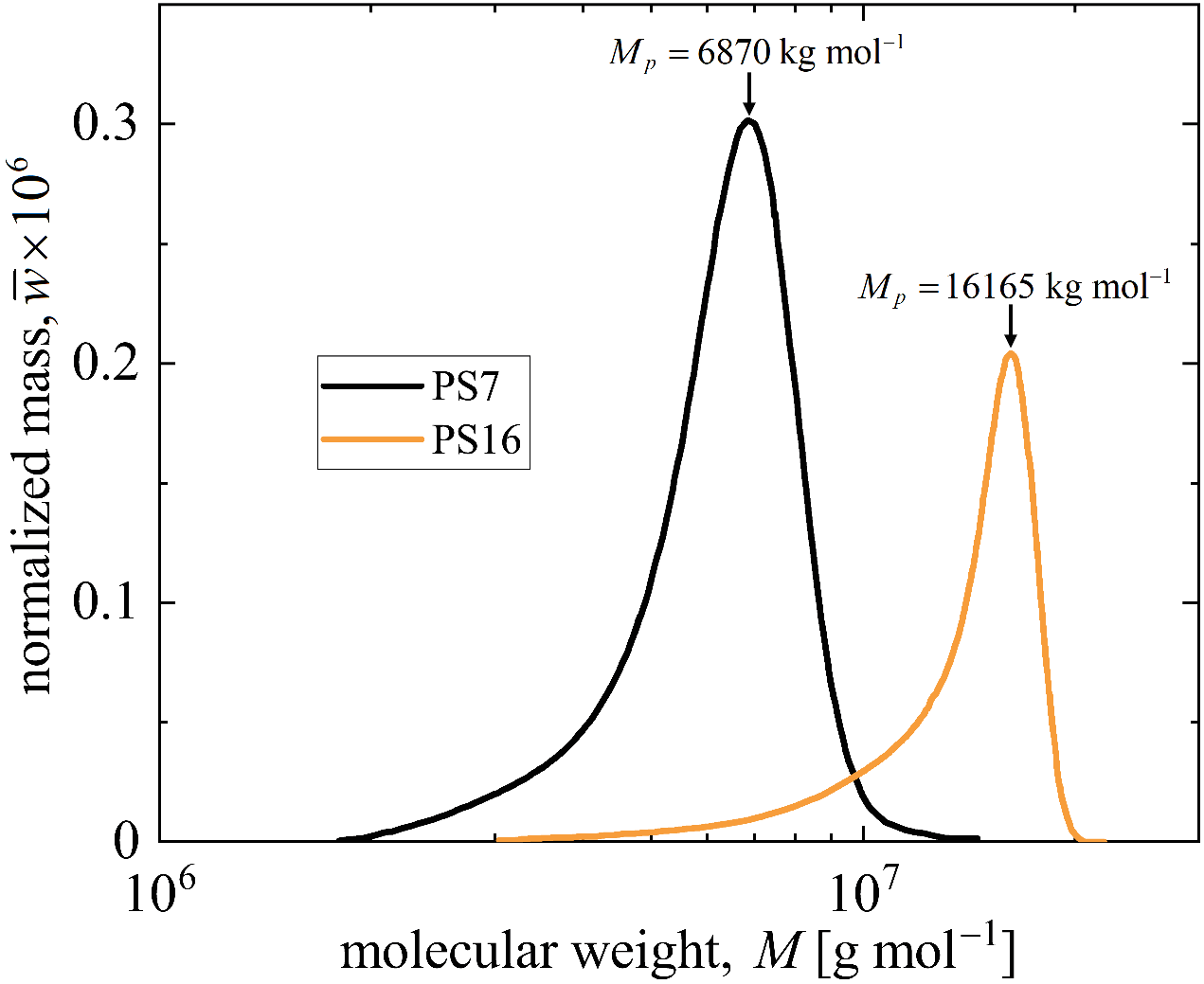}
    \caption{Normalized molecular weight distributions for the atactic polystyrene samples PS7 and PS16, from GPC analysis provided by the polymer supplier. 
 }
    \label{MWDs}
\end{figure}

Here, we propose that in fact MWD is a plausible explanation for the concentration-dependence of $\tau_{EC}$ at high dilutions, since the observation of an EC thinning regime depends on the generation of sufficient elastic stress by the stretching of polymer, which is itself concentration dependent \cite{Clasen2006}. Given a polymer sample with a distribution of molecular weights (two examples in Fig.~\ref{MWDs}), we argue that below a certain concentration there will be an insufficient number of molecules in the high-$M$ tail to generate the required elastic stress. In such a case, we imagine a scenario in which a lower-$M$ and more numerous portion of the MWD may dominate the EC response of the fluid. Since the relaxation time depends on the molecular weight $M$ of the polymer (as $\lambda \sim M^{3/2}$ for an ideal system), we propose that an interplay between the MWD and the polymer concentration dictates the determination of the elastocapillary timescale $\tau_{EC}$.

We elucidate the interplay between MWD and $c$ experimentally by employing two polymer samples with narrow and well-defined MWDs and distinct peak molecular weights $M_p$ (Fig.~\ref{MWDs}). We measure $\tau_{EC}$ for each of the samples individually over a range of concentration, and also for blends of the two samples, where the concentration of the low-$M_p$ component remains fixed and that of the high-$M_p$ component is progressively increased. For low concentrations of the high-$M_p$ component, the EC response is dominated by the low-$M_p$ component in the blend. However, if the concentration of the high-$M_p$ component in the blend is sufficient then it dominates the EC response and the low-$M_p$ component plays no apparent role. By assuming the progressive stretching of decreasing molecular weight species in the MWD and performing a backwards summation of the elastic stress contributions arising from each fully stretched molecular weight, the experimental results are qualitativley reproduced. Our findings are of significance to the application of capillary thinning techniques to extensional rheometry, to the interpretation of such measurements, and also to the understanding of elastocapillary thinning dynamics in general.

\begin{table*}[!ht]
\caption{Molecular parameters of the polystryrene samples under $\uptheta$-solvent conditions in DOP solution, evaluated for $M=M_p$.}
\label{Tab1}       
\center
\begin{ruledtabular}
\begin{tabular}{c c c c c c c c c c c c c c c}

polymer  &        $M_p$         &          $M_w$          & PDI            &  $N$   &   $b$  & $\langle R_0 \rangle$ & $R_g$  &       $l_c$        & $L$ &   $\lambda_Z$  & $c^*$ & $c_{min}$  & $c_{low}$  & $c^*_s$ \\
sample   & [kg~mol$^{-1}$] &  [kg~mol$^{-1}$]  &                   &           &  [nm] &             [nm]            &   [nm]   &  [$\upmu$m]  &       &       [ms]          & [ppm]  &      [ppm]    &      [ppm]  &    [ppm]  \\\noalign{\smallskip}\hline\noalign{\smallskip}
PS7        &        6870            &         6200            &    1.09         & 13600 &  1.49  &              174              &    71.2  &        20.3        & 117 &       1.03          &  3960  &       17.4      &     84        &    50   \\
PS16      &       16165           &        14480            &    1.07         & 32000 &  1.49  &              267              &   109    &        47.9        & 179 &       3.73          &  2580  &        4.8       &    44        &     21   \\

\end{tabular}
\end{ruledtabular}
\vspace{+0.1in}
\end{table*}

\section{Experimental methods}

\subsection{\label{Mat}Materials}

Two samples of high molecular weight and low polydispersity atactic polystyrene (PS) sourced from Agilent Technologies Inc. (Santa Clara, CA) are used in the experiments. One of the samples (labeled PS7) has a peak molecular weight $M_p = 6.87 \times 10^6~\text{g~mol}^{-1}$, with weight-averaged and number-averaged molecular weights of $M_w = 6.20 \times 10^6~\text{g~mol}^{-1}$ and $M_n = 5.67 \times 10^6~\text{g~mol}^{-1}$, respectively (polydispersity index, $\text{PDI}=M_w/M_n=1.09$). The other sample (labeled PS16) has $M_p = 16.17 \times 10^6~\text{g~mol}^{-1}$, $M_w = 14.48 \times 10^6~\text{g~mol}^{-1}$, and $M_n = 13.48 \times 10^6~\text{g~mol}^{-1}$ ($\text{PDI}=1.07$). The normalized MWDs of the two samples, obtained via gel permeation chromatography (GPC) characterization provided by the polymer supplier, are shown in Fig.~\ref{MWDs}. The samples have narrow distributions with clearly distinct peaks, but which overlap in a range of $3\times 10^6 \lesssim M \lesssim 14.5\times 10^6~\text{g~mol}^{-1}$.

Mother solutions of PS7 and PS16 are prepared at concentrations of $c_{\text{PS7}}=15,000~\text{ppm}$ and $c_{\text{PS16}}=10,000~\text{ppm}$ (by weight), respectively, in the solvent dioctyl phthalate (DOP) using the intermediate solvent method, as described in Ref.~\cite{Haward2016c}. 

Dioctyl phthalate is a moderately viscous ($\eta_s \approx 57~\text{mPa~s}$) and non-volatile $\uptheta$-solvent for polystyrene at $T=295~\text{K}$ (i.e., close to laboratory temperature) \cite{Berry1967}. Equilibrium chain parameters for PS in DOP can be estimated based on an ideal 3D random walk of Kuhn segments of length $b = C_{\infty}\ell $, and number of steps $N=n/C_{\infty}$ (neglecting bond angles), where $C_{\infty} = 9.7$ is the characteristic ratio, $\ell = 0.154~\text{nm}$ is the C-C bond length, $n=2M/m$ is the number of C-C bonds in a PS molecule of molar mass $M$, and $m=104~\text{g~mol}^{-1}$ is the molar mass of the styrene monomer \cite{Polymer_Handbook}. The equilibrium end-to-end length is $\langle R_0 \rangle = b \sqrt{N}$ and radius of gyration is $R_g = \langle R_0 \rangle/\sqrt{6}$. The overlap concentration can be estimated based on cubic packing of polymer coils as $c^* = M/(8N_AR_g^3)$ \cite{Graessley1980}. The contour length of the chain is given by $l_c = n\ell (\equiv Nb)$ and the polymer chain extensibility can be estimated as: 

\begin{equation}
L=l_c/ \langle R_0 \rangle.
\label{ext}
\end{equation}

Finally, the Zimm relaxation time for the ideal chain can be estimated as \cite{Zimm1956,Rubinstein_Colby}: 

\begin{equation}
\lambda_Z = \frac{1}{2\sqrt{3\uppi}} \frac{\eta_sR_g^3}{k_BT}.
\label{lam_Z}
\end{equation}

In order to observe a transition from Newtonian-like thinning to an exponential EC regime in a capillary thinning experiment the elastic stress due to the stretching polymer must exceed the viscous stress due to the Newtonian solvent. Clasen \textit{et al.} \cite{Clasen2006} have derived an expression for the minimum concentration $c_{min}$ of polymer which, when fully-stretched, would provide sufficient elastic stress to induce an observable EC regime:
\begin{equation}
c_{min} = \frac{3}{2} \frac{M \eta_s}{N_A k_B T \lambda_{Z} L^2},
\label{c_min}
\end{equation}
where $N_A$ is the Avogadro constant, $k_B$ is the Boltzmann constant, and $T$ is the absolute temperature.

Campo-Dea\~no and Clasen \cite{Campo2010} have derived another concentration criterion for EC thinning (termed $c_{low}$) describing the minimum concentration of polymer required to induce an EC regime at the onset of stretching (a supposed requirement for the extrapolation of $\tau_{EC} \equiv \lambda$ according to the prediction of the Oldroyd-B model, see Sec.~\ref{SRM}):

\begin{equation}
c_{low} = \frac{1}{2.46} \frac{M}{N_A k_B T} \left(\frac{\gamma^2\rho}{\lambda_{Z}^2}\right)^{1/3} \frac{1}{L^{3/2}},
\label{c_low}
\end{equation}
where $\gamma$ and $\rho$ are the surface tension and density of the fluid, respectively. For DOP at 25$^{\circ}$C, $\gamma \approx 0.031~\text{N~m}^{-1}$ and $\rho \approx 981~\text{kg~m}^{-3}$ \cite{Ricci1986}.

More recently, Dinic and Sharma \cite{Dinic2020} introduced a `stretched overlap concentration' $c^*_s$, which considers fully-stretched polymers as rod-like molecules and computes a corresponding overlap concentration based on the theory of Doi and Edwards \cite{Doi1986}:

\begin{equation}
c^*_s \approx \left(\frac{Nb}{dL^3}\right)c^*,
\label{c*_s}
\end{equation}
where $d$ is the diameter of a Kuhn segment and for PS $d \approx 1~\text{nm}$ \cite{Larson2005}. The value $c^*_s$ is proposed as an upper limit for ultradilution, above which concentration dependency of $\tau_{EC}$ is expected based on current arguments \cite{Dinic2020}. 

The values of all the molecular parameters and concentration criteria described above are presented for the PS7 and PS16 polymer samples (evaluated for $M=M_p$) in Table~\ref{Tab1}. The $c_{min}$, $c_{low}$, and $c^*_s$ criteria will be discussed in more detail later in the text.

Solutions of PS7 and PS16 are prepared at test concentrations $c_{\text{PS7}} = [200,~500,~1000]~\text{ppm}$ and $c_{\text{PS16}} = [5,~10,~20,~50,~100,~200]~\text{ppm}$ by progessive dilution of the mother fluids with additional DOP. The diluted fluids are placed on a roller mixer turning at 0.5~Hz until they are completely homogeneous (at least 24~hours) and subsequently left to stand and equilibrate for a further 24~hours prior to performing characterization tests or experimentation. As clear from the data provided in Table~\ref{Tab1}, the test concentrations of both PS7 and PS16 are all well below $c^*$, therefore can be considered dilute. Note also that all values of $c_{\text{PS7}}$ are significantly greater than either $c_{min,\text{PS7}}$ or $c_{low,\text{PS7}}$, so that at the tested concentrations all of the PS7 solutions could be expected to elicit a clear EC response during capillary thinning. On the other hand, for the PS16 solutions, as the concentration approaches $c_{\text{PS16}} \rightarrow c_{min,\text{PS16}}<c_{low,\text{PS16}}$, a diminishing EC response might be expected.


\begin{figure*}[ht!]
    \centering
    \includegraphics[width=17.0cm]{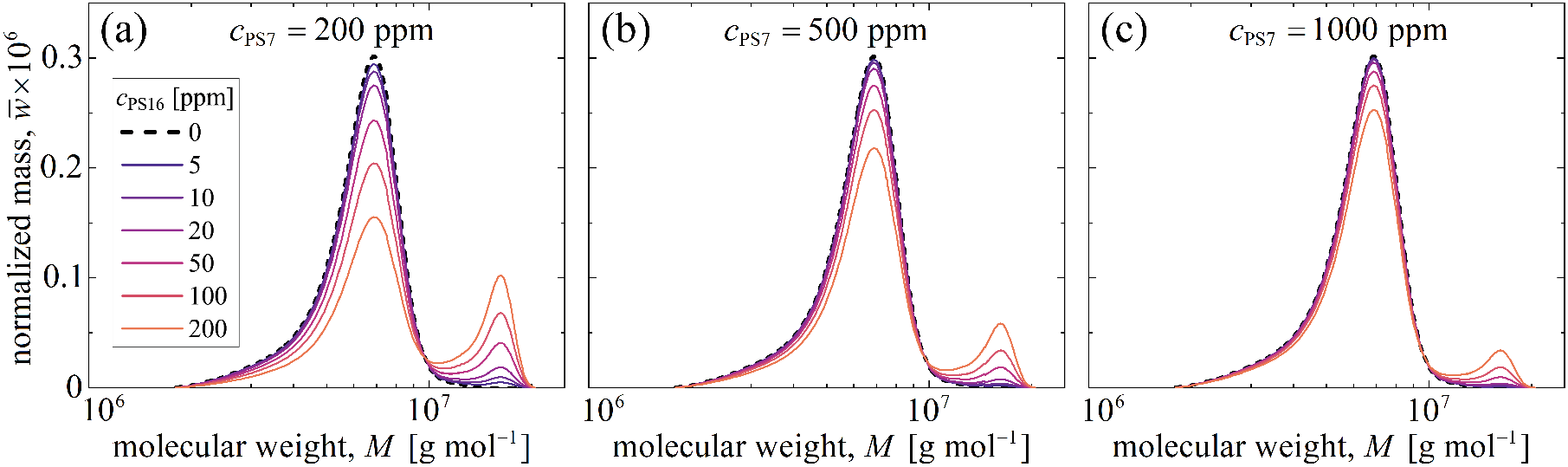}
\vspace{+0.1in}
    \caption{Normalized MWDs for blends of PS7 and PS16 at PS7 concentrations of (a) $c_{\text{PS7}} = 200~\text{ppm}$, (b) $c_{\text{PS7}} = 500~\text{ppm}$, (c) $c_{\text{PS7}} = 1000~\text{ppm}$, and at the PS16 concentrations $c_{\text{PS16}}$ indicated in the legend.
}
\vspace{+0.1in}
    \label{blendMWDs}
\end{figure*}

Blends at each concentration of $c_{\text{PS7}}$ and $c_{\text{PS16}}$ (total concentration $c = c_{\text{PS7}} + c_{\text{PS16}}$) are prepared by appropriate mixing of PS7 and PS16 solutions and the addition of further DOP, if necessary. As for the pure PS7 and PS16 test solutions, following the preparation of a blend the fluid is left on a roller mixer turning at 0.5~Hz until homogeneous and then allowed to stand for a further 24~hours prior to experimentation. Overlap concentrations for the blends can be estimated as $c^* \approx (c_{\text{PS7}} \times c^*_{\text{PS7}} + c_{\text{PS16}} \times c^*_{\text{PS16}})/c$. For the most concentrated mixture with $c_{\text{PS7}} = 1000~\text{ppm}$ and $c_{\text{PS16}} = 200~\text{ppm}$, the overlap concentration is $c^* \approx 3730~\text{ppm}$ (i.e., $c/c^* \approx 0.3$), indicating that all of the blended fluids remain in the dilute regime.

Combining the normalized MWDs shown in Fig.~\ref{MWDs} in appropriate ratios enables computation of the MWD of the polymer contained in each of the blended fluids. Normalized MWDs for the blends are shown in Fig.~\ref{blendMWDs}. For each concentration of PS7, the addition of a relatively small amount of PS16 essentially contributes a high-$M$ shoulder on the PS7 MWD. However, with increasing $c_{\text{PS16}}$, a clearly bimodal MWD emerges. It is important to note for interpretation of the results to be presented later that, due to the overlap between the PS7 and PS16 distributions (Fig.~\ref{MWDs}), the normalized mass of each blend is non-zero between the two peaks (i.e., around $M \approx 10^7~\text{g~mol}^{-1}$) and the mass in this region increases with increasing $c_{\text{PS16}}$.

All of the test fluids (i.e., pure PS7 and PS16 solutions and their blends) are characterized in steady shear at 25$^{\circ}$C using an Anton-Paar MCR502 stress-controlled rotational rheometer fitted with a 50~mm diameter 1$^{\circ}$ cone-and-plate geometry (see Fig.~S1, Supplementary Information Appendix). As expected for dilute polymer solutions, the fluids have viscosities $\eta$ close to that of the solvent $\eta_s \approx 57~\text{mPa~s}$ (the minimum solvent-to-total viscosity ratio is $\eta_s / \eta \approx 0.8$), and the fluids exhibit weak to negligible shear thinning.

\subsection{\label{SRM}Capillary thinning experiments}

When a liquid bridge confined between two interfaces becomes unstable to capillary forces (e.g., due to the separation between the interfaces being increased), surface tension acts to break the bridge and form two separate droplets (one on each confining interface) in order to minimize the total surface area of the fluid. Neglecting inertia, during the VC thinning prior to breakup of a viscous Newtonian fluid, the liquid bridge develops an hourglass shape with a minimum diameter $D$ at the midpoint (or neck) that reduces linearly in time $t$ \cite{Papageorgiou1995,McKinley2000}. The radial squeezing at the neck generates an axial extensional flow with a strain rate $\dot\varepsilon$ that increases with inverse proportionality to $D$ as $\dot\varepsilon = (-2/D)\text{d}D/\text{d}t$ \cite{McKinley2000,Anna2001b}, until the singular point of pinchoff where the fluid neck breaks and two separate droplets are formed. If polymer is added to the Newtonian fluid, the initial (inertialess) thinning behavior at early times is typically VC-like. However, as the extension rate in the fluid neck exceeds a critical value $\dot\varepsilon_c \approx 0.5/\lambda$ coiled-up polymer molecules unravel and stretch \cite{DeGennes1974,Hinch1977,McKinley2002}, leading to the generation of an entropic elastic stress that resists the capillary pressure and retards the thinning process. When a sufficient quantity of polymer has stretched such that the elastic stress exceeds the viscous stress, the VC balance is superceded by the EC balance \cite{Entov1997,Anna2001b,Clasen2006}. Now, when pinchoff would be expected for the Newtonian fluid, for the polymer solution the droplets that are formed on the two confining interfaces remain connected by a slender columnar filament. The diameter of the elastic filament is typically found to decay exponentially in time as $D \sim \exp(\frac{-t}{3\tau_{EC}})$, in accordance with the prediction of the Oldroyd-B model \cite{Bazilevsky1990,Entov1997,Anna2001b,McKinley2005}. Importantly, due to the exponential thinning in the EC regime, the extension rate in the filament becomes constant, with a value $\dot\varepsilon_{EC}=2/(3\tau_{EC})$ that is self-selected by the fluid via the stress balance \cite{Entov1997,Anna2001b,McKinley2005}. Several methods are utilized to generate and measure thinning capillary bridges for the purpose of finding $\tau_{EC}$ (and indeed other extensional rheological properties of complex fluids) \cite{Anna2001b,Campo2010,Nelson2011,Bhattacharjee2011,Dinic2015,Keshavarz2015,Sharma2015,Mathues2018,Rajesh2022,Gaillard2023}, with the most widely known and used being the capillary-breakup extensional rheometer (or CaBER) \cite{Anna2001b}. 

\begin{figure*}[ht!]
    \centering
    \includegraphics[width=17cm]{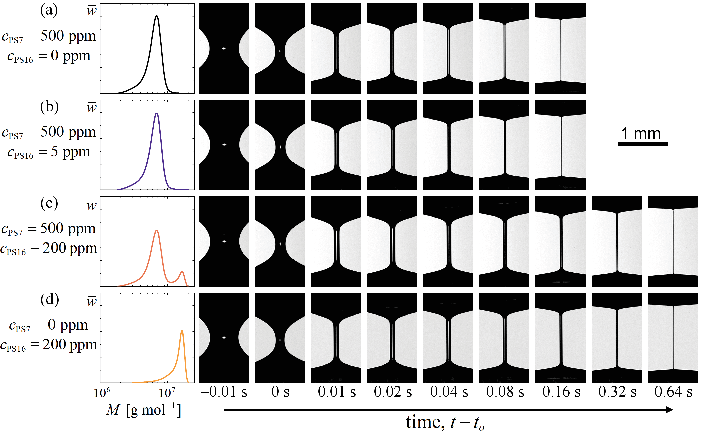}
    \caption{Imaging of the capillary thinning of a few exemplary polymer solutions characterized by the MWDs shown to the left: (a) 500~ppm of PS7 in pure solvent, (b) 500~ppm of PS7 plus 5~ppm of PS16, (c) 500~ppm of PS7 plus 200~ppm of PS16, (d) 200~ppm of PS16 in pure solvent. The time is indicated below the images where $t_o$ is the time at the onset of elastocapillarity. 
}
    \label{SRM_images}
\end{figure*}

In this work, the capillary thinning response of each test fluid is measured at 25$^{\circ}$C using the low-inertia slow retraction method (SRM \cite{Campo2010}) implemented on a commercial CaBER device (Haake CaBER 1, ThermoScientific). The test fluid is loaded such that it fills the $\approx 1~\text{mm}$ gap between the flat horizontal surfaces of two coaxial circular plates of diameter 6~mm. The lower plate, which is mounted on a micro-adjustable stage, is slowly displaced downwards increasing the plate separation and causing the fluid to neck and form a capillary bridge. When the gap between the plates is sufficiently large, the liquid bridge becomes unstable and begins to self-thin towards eventual pinchoff. The entire filament thinning and breakup process is captured at high speed (either 1000 or 3000~frames-per-second, depending on the total time until pinchoff) using a Phantom Miro M310 high speed video camera (Vision Research). A high intensity LED backlight (Fiber-Light, Dolan-Jenner) fitted with a telecentric lens (Edmund Optics) is used to render the image of the thinning liquid bridge in sharp sillouhette. The high speed camera is fitted with a $6.5 \times$ zoom lens (Edmund Optics) providing a resolution on the fluid filament of $\approx 5~\upmu \text{m~per~pixel}$, which is calibrated from an image of a 1~mm diameter wire. A Matlab program is used to process the resulting high frequency image sequences first by thresholding and then using edge detection to extract $D(t)$ at the midpoint between the two confining plates. The value of $\tau_{EC}$ is extracted in the exponentially decaying region of $D(t)$, as per the Oldroyd-B model. Measurements on each fluid are repeated five times with freshly loaded samples and with excellent reproducibility. Due to recent reports that the measurement of $\tau_{EC}$ by the SRM method can depend on the plate diameter \cite{Gaillard2023,Gaillard2024}, we also tested several fluids with plates of 4~mm and 8~mm diameter provided as standard with the commercial CaBER system. We found no systematic variation of the EC thinning dynamics over this range of plate size (see Fig.~S2, Supplementary Information Appendix).

\section{Results}
\vspace{-0.05in}
\subsection{Qualitative imaging of capillary thinning}
\vspace{-0.05in}

To highlight the fundamental phenomenon of interest that we report in this paper, and which we believe can explain the concentration dependence of $\tau_{EC}$, in Fig.~\ref{SRM_images} we present several exemplary time series of images depicting the capillary thinning of selected PS solutions. Here, the thinning of a 500~ppm solution of PS7 with no added PS16 (Fig.~\ref{SRM_images}(a)) appears to be minimally affected by the addition of 5~ppm of PS16 (Fig.~\ref{SRM_images}(b)). By contrast, the addition of 200~ppm of PS16 to a 500~ppm solution of PS7 (Fig.~\ref{SRM_images}(c)) results in a significantly ($\sim \mathcal{O}(4\times)$) longer time until pinchoff, and the thinning dynamics appear similar to those of a solution of pure 200~ppm PS16 with no PS7 in the mixture (Fig.~\ref{SRM_images}(d)).

Qualitatively it appears from the results shown in Fig.~\ref{SRM_images} that if $c_{\text{PS16}}$ is low, then the PS7 component dominates the capillary thinning response. However, if $c_{\text{PS16}}$ is high enough, then the PS16 component dominates the behavior. We therefore question the critical concentration conditions for the PS16 to dominate the dynamics, the understanding of which we anticipate should lead towards some fundamental insight into how polymer molecular weight and concentration mutually influence elastocapillary filament thinning.

\begin{figure*}[ht!]
    \centering
    \includegraphics[width=15cm]{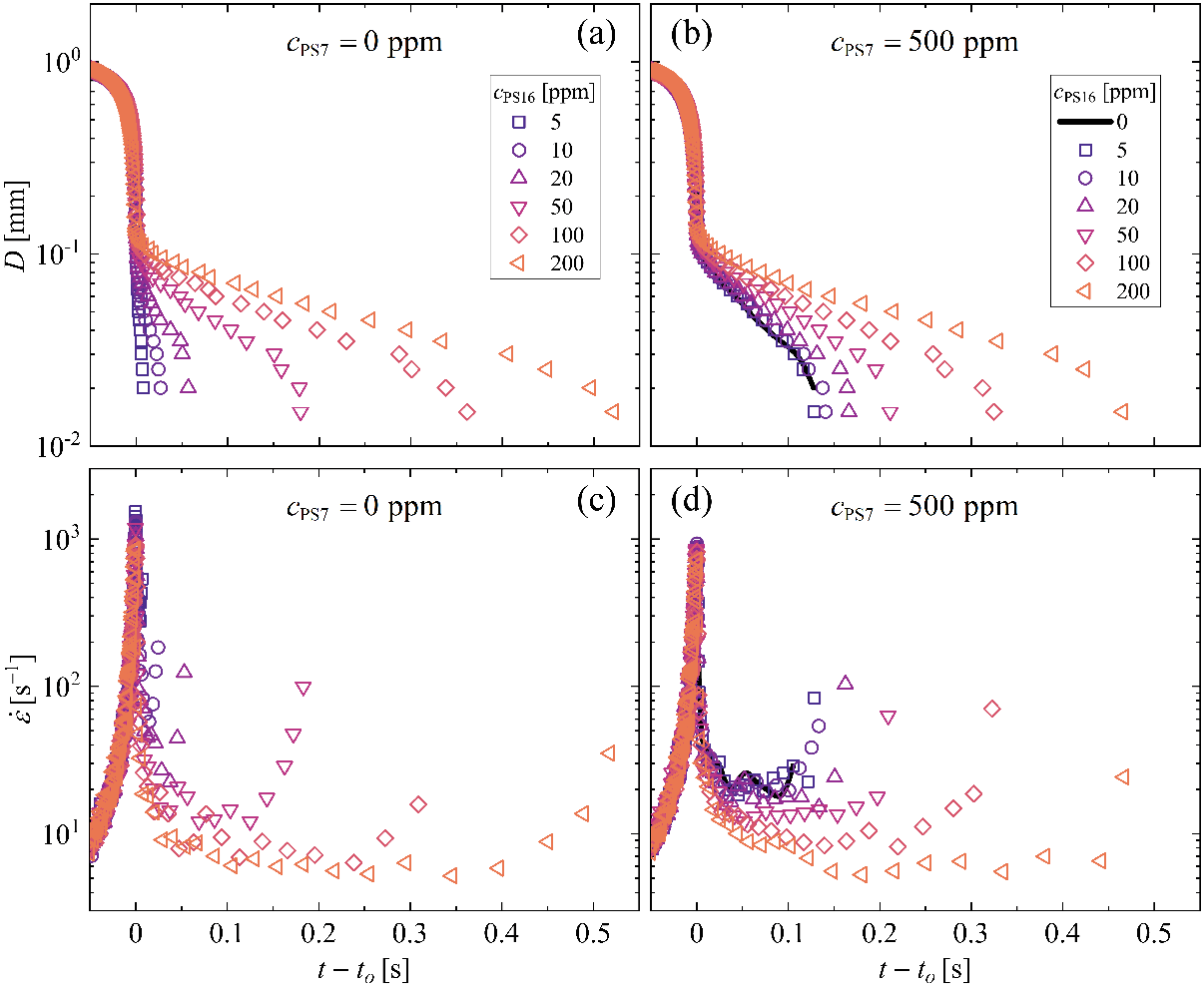}
    \caption{Comparison between the capillary thinning response of solutions of PS16 in pure solvent ($c_{\text{PS7}} = 0~\text{ppm}$) and mixtures of PS16 with 500~ppm of PS7. (a), (b) Filament diameter as a function of time for $c_{\text{PS7}} = 0~\text{ppm}$ and $c_{\text{PS7}} = 500~\text{ppm}$, respectively. (c), (d) Extension rate as a function of time for $c_{\text{PS7}} = 0~\text{ppm}$ and $c_{\text{PS7}} = 500~\text{ppm}$, respectively.
}
    \label{SRM_profiles}
\end{figure*}

\subsection{Capillary thinning dynamics}

Important detail on how the capillary thinning dynamics depend on the fluid composition can be elucidated by plotting filament diameter \emph{versus} time curves such as those shown in Fig.~\ref{SRM_profiles}(a,b). Note that in these plots the time axis is shifted by the value $t_o$ marking the onset of the EC thinning regime. Time $t_o$ occurs at the peak in the respective extension rate \emph{versus} time plot (see Fig.~\ref{SRM_profiles}(c,d)). In Fig.~\ref{SRM_profiles}(a) we present filament thinning profiles for pure PS16 solutions (i.e., $c_{\text{PS7}} = 0~\text{ppm}$) over the full range of $c_{\text{PS16}}$. At $c_{\text{PS16}}=5~\text{ppm}$ the breakup time is very short ($\sim \mathcal{O}(\text{ms})$) and the thinning in the EC regime ($t > t_o$) is relatively fast. With increasing $c_{\text{PS16}}$, the thinning rate in the EC regime progressively decreases and the time to pinchoff progressively increases. When the same concentrations of PS16 are blended with 500~ppm of PS7, the thinning curves are markedly shifted, at least for low values of $c_{\text{PS16}}$. For $c_{\text{PS16}} \leq 10~\text{ppm}$ the EC thinning in combination with 500~ppm of PS7 (Fig.~\ref{SRM_profiles}(b)) is significantly slower than it is in the presence of 0~ppm of PS7 (Fig.~\ref{SRM_profiles}(a)). At these low concentrations, the presence of PS16 appears to have little effect on the thinning dynamics compared to the background 500~ppm PS7 solution (shown by the solid black line in Fig.~\ref{SRM_profiles}(b)). However, at higher concentrations of $c_{\text{PS16}} \geq 20~\text{ppm}$ there is a noticable reduction in the thinning rate and increase in the time to pinchoff. For $c_{\text{PS16}} \geq 100~\text{ppm}$, a comparison between the thinning profiles shown in Fig.~\ref{SRM_profiles}(a) and Fig.~\ref{SRM_profiles}(b) reveals that the response is largely unaffected by the presence (or not) of PS7 in the blend. 

The plots of extension rate \emph{versus} time provided in Fig.~\ref{SRM_profiles}(c,d) show how $\dot\varepsilon$ increases monotonically during the Newtonian-like (or VC) thinning regime ($t < t_o$), reaching a peak at $t_o$. The time $t_o$ marks the instant at which the elastic stress due to the stretching of dissolved polymer begins to exceed the viscous stress due to the solvent, i.e., the onset of the EC thinning regime. Subsequently, for $t-t_o > 0$, the extension rate decreases to the plateau value $\dot\varepsilon_{EC}=2/(3\tau_{EC})$. It is worthwhile noting that the value of $\dot\varepsilon_{EC}$ progressively decreases for increasing $c_{\text{PS16}}$ for both cases of $c_{\text{PS7}} = 0~\text{ppm}$ (Fig.~\ref{SRM_profiles}(c)) and $c_{\text{PS7}} = 500~\text{ppm}$ (Fig.~\ref{SRM_profiles}(d)).

\begin{figure}
    \centering
    \includegraphics[width=8.0cm]{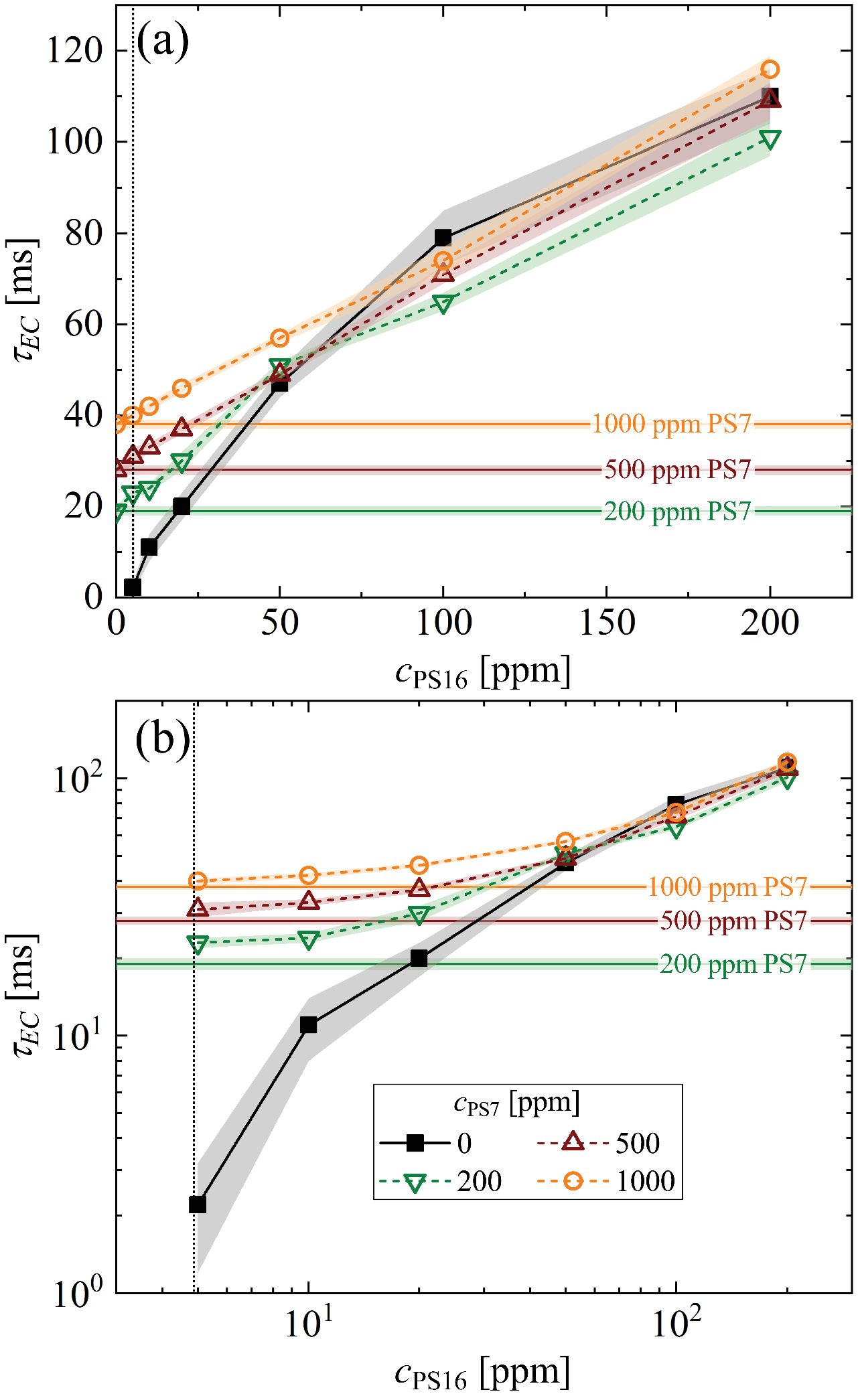}
    \caption{Values of $\tau_{EC}$ determined for all tested polymer solutions as a function of the PS16 concentration. Thin horizontal solid lines indicate $\tau_{EC}$ for the respective PS7 concentration and $c_{\text{PS16}} = 0~\text{ppm}$. Shaded regions about the data indicate one standard deviation over five repeated measurements. The vertical dotted line marks the value of $c_{min,{\text{PS16}}} = 4.8~\text{ppm}$ Parts (a) and (b) show the same data but on linear-linear and log-log scales, respectively.
 }
    \label{tauEC}
\end{figure}

\subsection{Characteristic EC thinning times}

Characteristic times $\tau_{EC}$ extracted in the EC regime for all of the tested polymer solutions are displayed in Fig.~\ref{tauEC}, where parts (a) and (b) report the same data but presented on linear-linear and log-log axes, respectively. In these plots, the solid horizontal lines mark the values of $\tau_{EC}$ determined for pure PS7 solution (i.e., with $c_{\text{PS16}} = 0~\text{ppm}$), and solid black data points connected by solid lines mark the values of $\tau_{EC}$ determined for pure PS16 solutions (i.e., with $c_{\text{PS7}} = 0~\text{ppm}$). The open colored data points connected by dashed lines mark $\tau_{EC}$ for blends of PS7 and PS16, with $c_{\text{PS7}}$ indicated in the legend, and $c_{\text{PS16}}$ indicated on the horizontal axis. Shaded regions about the data points and/or lines indicate the standard deviation over five repeated measurements of $\tau_{EC}$. 

For pure solutions of PS16, $\tau_{EC}$ increases monotonically with increasing $c_{\text{PS16}}$. For $c_{\text{PS16}} = 5~\text{ppm}$ (which is only slightly above $c_{min,\text{PS16}} = 4.8~\text{ppm}$, marked by the vertical dotted line), $\tau_{EC}$ is difficult to measure at only $\approx 2~\text{ms}$ and it appears would become immeasurable at lower concentrations. However, we note that the apparent agreement between our measurement limit and $c_{min,\text{PS16}}$ (Eq.~\ref{c_min}) is probably coincidental since $c_{min,\text{PS16}}$ is estimated for $M_p$ and in no way accounts for the MWD. We will return to this point later in the text. Notice that although our data is quite coarsely spaced in concentration, we see no tendency to a constant value of $\tau_{EC}$ for low values of $c_{\text{PS16}}<c^*_s\approx20~\text{ppm}$. There is perhaps a tendency towards a high concentration plateau value of $\tau_{EC}$ omewhat greater than $\approx 100~\text{ms}$.  

For the polymer solution blends with low $c_{\text{PS16}} \lesssim 10~\text{ppm}$, the measured value for $\tau_{EC}$ is close to, but slightly greater than, that measured for $c_{\text{PS16}} = 0~\text{ppm}$ (i.e., for the respective pure PS7 solution). However, with increasing $c_{\text{PS16}} \gtrsim 50~\text{ppm}$, $\tau_{EC}$ for all of the blends (regardless of PS7 concentration) tends closely to that of the pure PS16 solution (i.e., with $c_{\text{PS7}} = 0~\text{ppm}$).

\subsection{Interpretation of the results}

We interpret the data summarized in Figs.~\ref{SRM_profiles}~and~\ref{tauEC} as described in the following. We know with confidence that (1) the relaxation time of the polymer depends on molecular weight as $\lambda \sim M^{3/2}$ \cite{Keller1985,Rabin1985}, (2) polymer stretching in the flow is predicted for $\dot\varepsilon_c \sim 1/\lambda$ \cite{DeGennes1974,Hinch1977,Perkins1997} and (3) in the VC regime the extension rate $\dot\varepsilon$ increases with time (as shown by Fig.~\ref{SRM_profiles}(c,d) for $t<t_o$). Therefore we expect that as time proceeds through the VC regime, and $\dot\varepsilon$ increases, progressively shorter, lower-$M$ molecules in the MWD will begin to stretch. We assume that eventually, at time $t = t_o$, a sufficient fraction of the MWD will have become stretched such that the total accumulated elastic stress exceeds the viscous stress, therefore resulting in the onset of the EC regime. 

\begin{figure}[ht!]
    \centering
    \includegraphics[width=8cm]{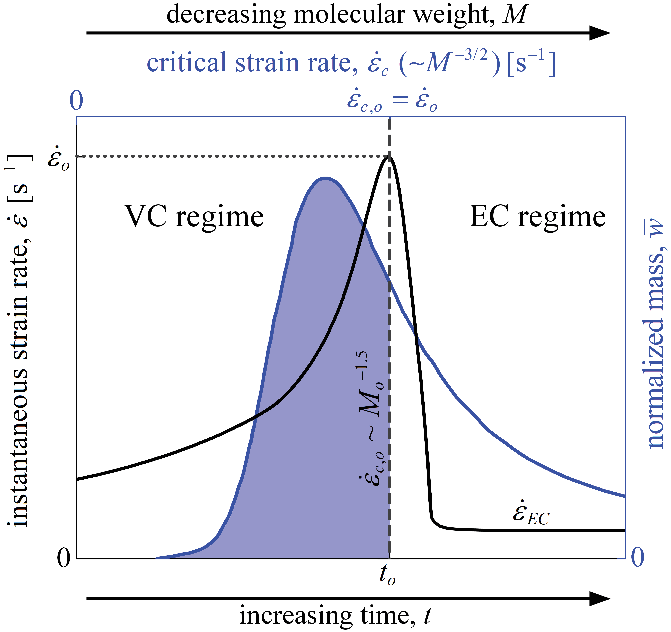}
    \caption{Sketch outline of the proposed hypothesis involving the stretching of a portion of the MWD distribution during the VC thinning regime. The solid black line represents an idealized extension rate \emph{versus} time curve in a self-thinning capillary neck. The solid blue line represents a distribution of critical extension rates $\dot\varepsilon_c$ for the stretching of the polymers in an hypothetical MWD. The horizontal dotted line marks the peak extension rate $\dot\varepsilon_o$ at time $t=t_o$ marking the VC-EC transition. The vertical dashed line marks time $t_o$, and also the critical extension rate $\dot\varepsilon_{c,o} (\equiv \dot\varepsilon_o)$ of the shortest polymers to stretch (with $M=M_o$) before the onset of the EC regime. The shaded region under the blue curve indicates the portion of the MWD with $M > M_o$ that should have already stretched (or started to stretch) prior to $t_o$.
 }
    \label{cartoon}
\end{figure}

This concept is outlined in Fig.~\ref{cartoon}, where we schematically represent an idealized extension rate \emph{versus} time curve for a thinning polymer solution overlaid on a distribution of critical strain rates $\dot\varepsilon_c(M) \sim 1/M^{3/2}$ corresponding to a hypothetical MWD. Fig.~\ref{cartoon} illustrates that (in principle) the peak extension rate $\dot\varepsilon_o$ achieved at time $t_o$ corresponds to the critical extension rate $\dot\varepsilon_{c,o}$ of the shortest polymers to stretch (with molecular weight $M_o$) prior to the onset of the EC regime. All polymers with $M>M_o$ (thus $\dot\varepsilon_{c} < \dot\varepsilon_{c,o}$, shaded region) are assumed to have already stretched at earlier times $t<t_o$. The total polymer concentration illustrated in Fig.~\ref{cartoon} is arbitrary. For increasing (decreasing) polymer concentrations, $t_o$ is expected to shift left (right) since less (more) of the MWD is required to stretch in order to generate sufficient elastic stress for an EC regime to manifest. Correspondingly, $\dot\varepsilon_o$ is expected to decrease (increase) in order to match $\dot\varepsilon_{c,o}$. The expected variation of $\dot\varepsilon_o$ with polymer concentration is observed experimentally (e.g., Fig.~\ref{SRM_profiles}(c) and Fig.~S3, Supplementary Information Appendix), and was also reported in Ref. \cite{Rajesh2022}. However, we find considerable uncertainty in the extraction of $\dot\varepsilon_o$ since it is represented by a single data point computed by differentiation of noisy raw data.

Returning to the experimental results in Figs.~\ref{SRM_profiles}~and~\ref{tauEC}, we suggest that for the pure PS16 solutions if $c_{\text{PS16}} \lesssim 5~\text{ppm}$ then the whole of the MWD can stretch without generating enough elastic stress to induce an EC regime. If $c_{\text{PS16}}$ is small but $\gtrsim 5~\text{ppm}$ then we suggest that most of the PS16 MWD must stretch prior to the onset of the EC regime, thus the EC regime is controlled by a low-$M$ portion of the MWD. As $c_{\text{PS16}}$ is increased, a smaller fraction of the MWD is required to stretch before the elastic stress exceeds the viscous stress. Thus, the EC regime is achieved within a higher molecular weight portion of the distribution, the longer relaxation time of which results in a larger value for $\tau_{EC}$.

In terms of the polymer blends, even the smallest concentration of PS16 will contribute some elastic stress when deformed, thus $\tau_{EC}$ measured for the blend is always greater than that measured for the respective pure PS7 solution. Recall that the tails of the PS7 and PS16 MWDs have an overlapping region (see Fig.~\ref{MWDs}). Accordingly, although a pure PS16 solution at $c_{\text{PS16}} = 5~\text{ppm}$ can by itself induce an EC regime, it is still possible for part of the high-$M$ tail of the PS7 distribution to stretch and contribute to the total elastic stress. At higher values of PS16 (i.e., $c_{\text{PS16}} \gtrsim 50~\text{ppm}$) we suggest that sufficient elastic stress is derived from the high-$M$ tail of the PS16 MWD (at molecular weights beyond the range of overlap) so that none of the PS7 MWD is required (or able) to deform. Accordingly, at these high PS16 concentrations, the PS7 plays essentially no role in the emerging filament thinning dynamics and $\tau_{EC}$ for the blend closely matches that for the respective pure PS16 solution.

\begin{figure*}[ht!]
    \centering
    \includegraphics[width=18cm]{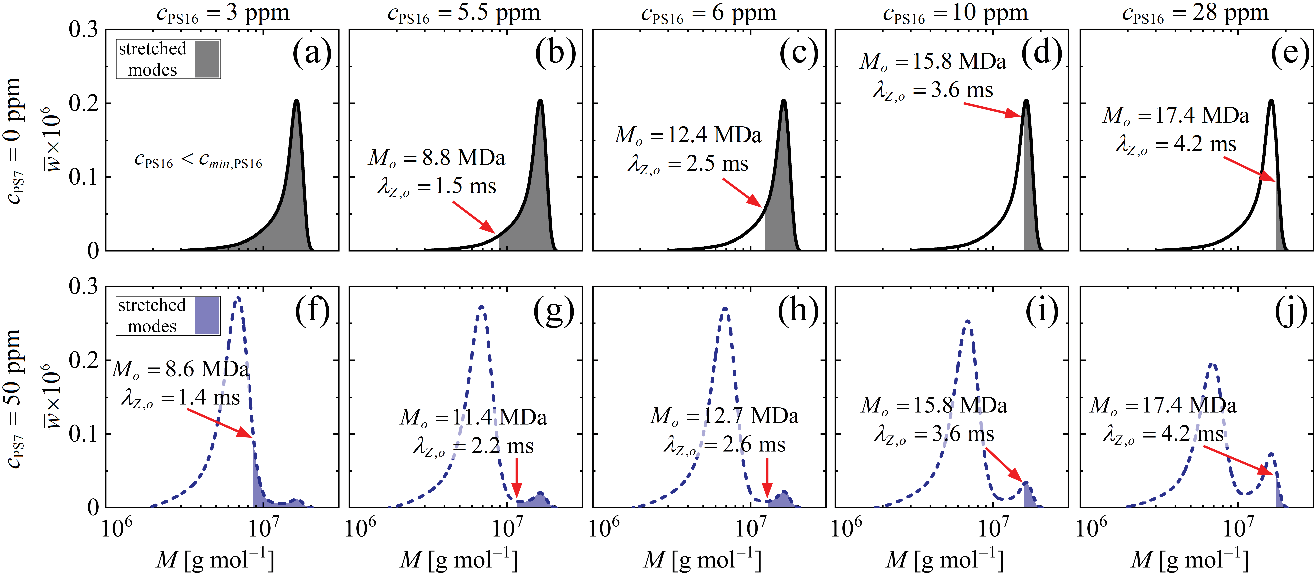}
    \caption{Values of the Zimm relaxation time $\lambda_{Z,o}$ corresponding to the lowest molecular weight mode $M_o$ at which the accumulated elastic stress exceeds the viscous stress, computed by summation of the elastic stress contributions of each fully-stretched mode (see Eq.~\ref{stress} and text description). Shaded regions underneath the curves indicate the portion of the MWD that must become fully-stretched during the VC thinning regime in order to induce a VC-EC transition. Parts (a-e) are obtained for MWDs with $c_\text{PS7} = 0~\text{ppm}$, and parts (f-j) for $c_\text{PS7} = 50~\text{ppm}$, with corresponding values for $c_\text{PS16}$ indicated at the top of the figure. 
 }
    \label{lambdaZ_comp}
\end{figure*}

\subsection{A simple model description}

Our hypothetical explanation for our results suggests a relatively simple qualitative description. Closely following the procedure used in the derivation of the $c_{min}$ criterion (Eq.~\ref{c_min}) \cite{Clasen2006}, we aim to compute and sum the elastic stress contributions arising from the progressive stretching of decreasing molecular weight species in a distribution. We bin the normalized MWDs such as illustrated in Figs.~\ref{MWDs}~and~\ref{blendMWDs} into ``modes'' $i$ each spanning a range of $M=20,000~\text{g~mol}^{-1}$, where mode $i=1$ has the highest $M$, mode $i=2$ the second-highest $M$, \emph{etc}. For a given total polymer concentration $c = c_{\text{PS7}} + c_\text{{PS16}}$, we use the normalized MWD to compute the concentration $c_i$ of polymer contained in each mode. For each mode with molecular weight $M_i$, we compute: (1) the polymer extensibility $L_i$ (according to Eq.~\ref{ext}), (2) the Zimm relaxation time $\lambda_{Z,i}$ (according to Eq.~\ref{lam_Z}), and (3) the elastic modulus $G_i = c_i N_A k_B T/M_i$ \cite{Ferry1980,Clasen2006}. The elastic stress contributed by each mode when fully stretched is estimated as \cite{Clasen2006}: 

\begin{equation}
\Delta \sigma_{p,i} = 2G_i \lambda_{Z,i} L_i^2 \dot\varepsilon .
\label{modestress}
\end{equation}
Note that at a given strain rate it takes finite time for macromolecular strain to accumulate and so it is unreasonable to expect every stretching mode to attain full extension within time $t \leq t_o$. However, Eq.~\ref{modestress} should provide an estimated upper-bound for the stress contributed by each mode. 

The viscous stress due to the solvent is given simply by $\Delta \sigma_{s} = 3\eta_s \dot\varepsilon$, where the factor $3$ represents the Trouton ratio in uniaxial extensional flow. 

We sum the elastic stress contributions from progressively shorter modes, until the total exceeds the viscous stress, i.e.:

 \begin{equation}
\frac{2}{3\eta_s} \sum_{i=1}^{i=o} G_i \lambda_{Z,i} L_i^2 > 1.
\label{stress}
\end{equation}
 
Finally, we extract the properties of the last (shortest) mode in the summation, specifically $M_o$ and $\lambda_{Z,o}$. This process is approximately predicting the stretched portion of the MWD with $M \geq M_o$ at the onset of the EC regime at $t=t_o$. We are not attempting to account for any relaxation of shorter modes as the extension rate decreases towards the plateau value $\dot\varepsilon_{EC}$ subsequent to $t_o$. Implicit assumptions here are that molecules are completely non-interacting and that all modes adopt the Zimm relaxation time regardless of the input total polymer concentration. We also point out that the Zimm relaxation times refers to the time for a coiled non-free draining polymer molecule to diffuse through the solvent a distance equal to its own size and does not necessarily describe a timescale related with stretching and/or recoil of the chain \cite{Rubinstein_Colby}. Therefore we do not necessarily expect to obtain a quantitative agreement between $\lambda_{Z,o}$ and our measured values for $\tau_{EC}$. However, $\lambda_{Z,i}$ is conveniently calculable and does give the expected molecular weight scaling for stretching timescales of ideal polymers in extensional flows \cite{Farrell1980,Fuller1980,Keller1985,Rabin1985,Rabin1985b}.

Fig.~\ref{lambdaZ_comp}(a-e) illustrates the effect of varying the total polymer concentration $c$ on the portion of the PS16 MWD that must stretch in order to induce an EC regime, where the stretched modes are indicated by shading under the MWD curve. Here, since $c_\text{PS7} = 0~\text{ppm}$, then $c \equiv c_\text{PS16}$. At very low $c_\text{PS16} =3~\text{ppm}$ (Fig.~\ref{lambdaZ_comp}(a)) even when the whole MWD is fully stretched there is insufficient elastic stress to overcome the viscous stress, hence we predict no EC regime and we cannot obtain values for $M_o$ or $\lambda_{Z,o}$ in this case. The elastic stress contributed by the whole of the fully-stretched PS16 MWD remains below the viscous stress as long as $c_{\text{PS16}} \lesssim 5.4~\text{ppm}$. We can define the concentration $c'_{min,\text{PS16}} \approx 5.4~\text{ppm}$ as a measure of the $c_{min}$ criterion (Eq.~\ref{c_min} \cite{Clasen2006}) that accounts for the entire PS16 MWD (though which in this case happens to be coincidentally close to the value $c_{min,\text{PS16}}= 4.8~\text{ppm}$ estimated based on a single molecular weight $M_p$). For $c_\text{PS16} =5.5~\text{ppm}~(\gtrsim c'_{min,\text{PS16}}$, Fig.~\ref{lambdaZ_comp}(b)), most of the PS16 distribution must stretch in order to generate the required elastic stress, hence $M_o$ is located in the low-$M$ tail of the MWD with $\lambda_{Z,o} = 1.5~\text{ms}$. With increasing $c_\text{PS16}$ through Fig.~\ref{lambdaZ_comp}(c-e), progressively less of the MWD is required to stretch in order for the viscous stress to be exceeded, thus $M_o$ shifts towards the high-$M$ tail of the MWD, and $\lambda_{Z,o}$ progressively increases. 

\begin{figure*}[ht!]
    \centering
    \includegraphics[width=15.0cm]{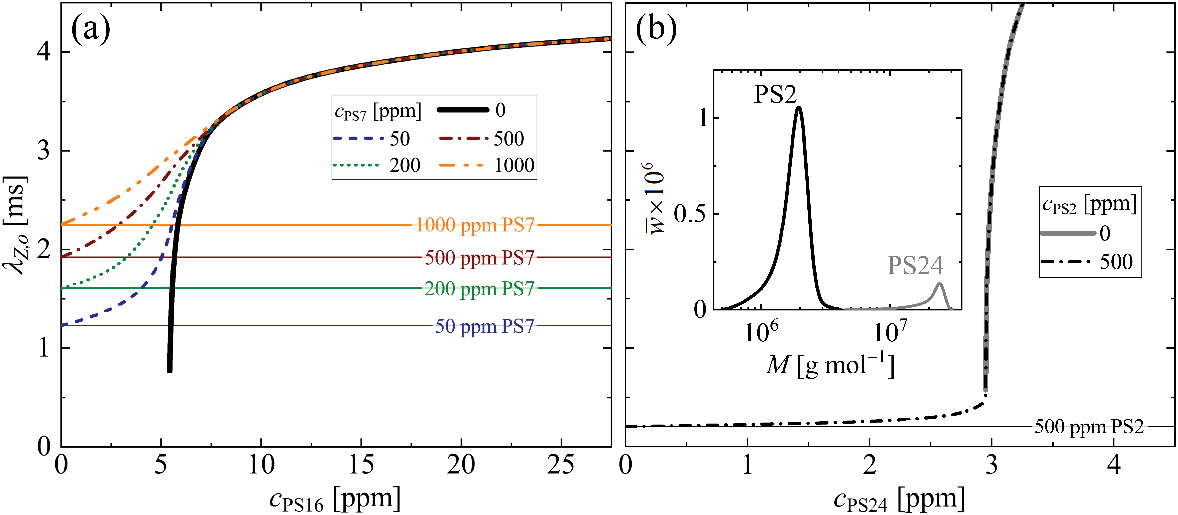}
    \caption{Values of the Zimm relaxation time of the last fully stretched mode $\lambda_{Z,o}$ at the point where the elastic stress exceeds the viscous stress, computed according to Eq.~\ref{stress}. (a) Data determined using the overlapping MWDs for PS7 and PS16 (see Fig.~\ref{MWDs}) with various values of $c_\text{PS7}$ and for refined increments in $c_\text{PS16}$, (b) data determined using the synthetic non-overlapping MWDs for PS2 and PS24 (as shown in the insert plot), with $c_\text{PS2} = 500~\text{ppm}$ and for refined increments in $c_\text{PS24}$. 
 }
    \label{lambdaZ_comp2}
\end{figure*}

Fig.~\ref{lambdaZ_comp}(f-j) illustrates the effect of varying $c_\text{PS16}$ in a blend with $c_\text{PS7} = 50~\text{ppm}$ (i.e., such that the total polymer concentration $c = c_\text{PS16} + 50~\text{ppm}$). At $c_\text{PS16} =3~\text{ppm}$, whereas the pure PS16 solution (Fig.~\ref{lambdaZ_comp}(a)) could not yield values for $M_o$ or $\lambda_{Z,o}$, in the corresponding blend (Fig.~\ref{lambdaZ_comp}(f)) the stretching of the high-$M$ tail of the PS7 MWD leads to sufficient elastic stress to induce the EC regime at $M_o = 8.6 \times 10^6~\text{g~mol}^{-1}$ with $\lambda_{Z,o} = 1.4~\text{ms}$. As observed for pure PS16 solutions, increasing $c_\text{PS16}$ in the blend (thus increasing $c$) causes $M_o$ to shift to progressively higher values. At $c_\text{PS16} = 5.5~\text{ppm}$ and $c_\text{PS16} = 6~\text{ppm}$, $M_o$ for the pure PS16 solution occurs within the range where the PS7 and PS16 MWDs overlap (see Fig.~\ref{lambdaZ_comp}(b,c) and Fig.~\ref{MWDs}). Thus, at these values of $c_\text{PS16}$ in the blend (see Fig.~\ref{lambdaZ_comp}(g,h)), $M_o$ is greater (and $\lambda_{Z,o}$ higher) than that of the pure PS16 solution due to the increased concentration of polymer $c_i$ in the overlapping modes. As $c_\text{PS16}$ is further increased to 10~ppm and above, $M_o$ for the pure PS16 solution shifts beyond the range of overlap between the two MWDs. In these cases $M_o$ for the blend (Fig.~\ref{lambdaZ_comp}(i,j)) becomes identical to that of the pure PS16 (Fig.~\ref{lambdaZ_comp}(d,e)); i.e., sufficient elastic stress is developed by the high-$M$ portion of the PS16 MWD such that, in the blend, none of the PS7 MWD is required to stretch. In a real capillary thinning experiment, this situation would correspond to a case in which the extension rate prior to the EC regime never became high enough to stretch any of the molecules in the PS7 MWD.

Note that here we have used a blend with lower $c_\text{PS7}$ than used experimentally simply to permit a slightly clearer visualization of the small amount of PS16 in the blend. However, precisely the same principles apply for blends with higher PS7 concentrations.

In Fig.~\ref{lambdaZ_comp2}(a) we present computed values of $\lambda_{Z,o}$ as a function of $c_\text{PS16}$ for various blends at different values of $c_\text{PS7}$, mimicking the presentation of the experimental data in Fig.~\ref{tauEC}(a). Calculations are performed at sufficiently small increments in $c_\text{PS16}$ to generate effectively continuous line plots. The simple summation of elastic stresses performed to obtain $M_o$ and $\lambda_{Z,o}$ (Eq.~\ref{stress}) helps to corroborate the hypothesis derived on the basis of our experiments. In Fig.~\ref{lambdaZ_comp2}(a) our calculations show that for pure PS16 solutions, for $c>c'_{min,\text{PS16}}\approx5.4~\text{ppm}$ the relaxation time increases abruptly and monotonically. Relaxation times of the PS7 and PS16 blends are close to (but always greater than) that of the pure PS7 solution for low $c_\text{PS16}<c'_{min,\text{PS16}}$ since even at vanishing concentrations stretching of the PS16 contributes finite elastic stress. With increasing $c_\text{PS16}<c'_{min,\text{PS16}}$, the relaxation time of the blend diverges from that of the respective pure PS7 solution due to the progressively increasing elastic stress contributed by stretched modes in the PS16 MWD. At intermediate PS16 concentrations $c'_{min,\text{PS16}} < c_\text{PS16} \lesssim 8~\text{ppm}$, even though the PS16 MWD alone can generate sufficient elastic stress to provide a value for $\lambda_{Z,o}$, the relaxation times for the blends remain greater than for the pure PS16. This is due to the overlap between the PS7 and PS16 MWDs and the increased concentration of material in the shared modes. At higher values of $c_\text{PS16} \gtrsim 8~\text{ppm}$, the relaxation time of each blend matches that of the pure PS16 since sufficient stress is generated by the stretching of the highest-$M$ modes in the PS16 MWD that none of the modes within the PS7 MWD are required to stretch. Given the significant simplifying assumptions underlying our retrieval of $\lambda_{Z,o}$ by Eq.~\ref{stress}, the clear qualitative agreement between Fig.~\ref{lambdaZ_comp2}(a) and Fig.~\ref{tauEC}(a) is remarkable and compelling, with all the essential trends of the experimental data being captured by our simple model.

The overlap between the PS7 and PS16 MWDs complicates our analysis somewhat, but we are also able to use our model to examine the response of blends whose MWDs do not overlap at the extremes. Unfortunately, this is complicated experimentally due to the very long tails of real MWDs and because the short relaxation times of low-$M$ polymers are hard to measure reliably. For these reasons, obtaining two polymer samples, both with measureable $\tau_{EC}$ at dilute concentrations, and with sufficiently widely separated values of $M_p$ is challenging. In the case of applying our simple model to the problem, non-interacting MWDs are easily generated by shifting the PS7 MWD by a factor $1/3.5$ and by shifting the PS16 MWD by a factor $1.5$ along the $M$-axis, thus providing synthetic `PS2' and `PS24' MWDs, as illustrated in the insert to Fig.~\ref{lambdaZ_comp2}(b). The main plot in Fig.~\ref{lambdaZ_comp2}(b) shows values of $\lambda_{Z,o}$ computed for pure PS24 solutions, and for blends of 500~ppm PS2 with PS24, over a range of $c_\text{PS24}$. For the pure PS24 solutions, we cannot determine a value for $\lambda_{Z,o}$ if $c_{\text{PS24}} \lesssim 3~\text{ppm}$ since below this concentration the elastic stress is below the viscous stress even if the whole of the PS24 MWD is fully-stretched. Hence we can define $c'_{min,{\text{PS24}}} \approx 3~\text{ppm}$. For increasing concentrations of $c_{\text{PS24}} > 3~\text{ppm}$, $\lambda_{Z,o}$ for the pure PS24 solutions increases abruptly. For the blends with 500~ppm PS2 and $0 < c_{\text{PS24}} \lesssim 3~\text{ppm}$,  $\lambda_{Z,o}$ \emph{can} be determined and is close to that of the pure 500~ppm PS2 solution. In this range of PS24 concentration, $\lambda_{Z,o}$ increases slightly with increasing $c_{\text{PS24}}$ due to the small, but increasing, elastic stress contribution arising from the full stretching of the entire PS24 MWD, which shifts $M_o$ progressively towards the high-$M$ tail of the PS2 MWD. Since there is now no cross-coupling between extreme modes in the two MWDs, for $c_{\text{PS24}} \gtrsim 3~\text{ppm}$ the PS24 dominates the response of the blend, with the PS2 MWD contributing precisely zero elastic stress in our simple model. Thus, for $c_{\text{PS24}} \gtrsim 3~\text{ppm}$, $\lambda_{Z,o}$ for the blend precisely matches that for the pure PS24 solution.

\section{Discussion}

With a view to obtaining new insight into the importance of the molecular weight distribution of real polymer samples on EC thinning measurements with dilute polymer solutions, we have performed experiments with blends of ideal and well-characterized polymer samples having distinct peak molecular weights $M_p$ and narrow MWDs. By fixing the concentration of the lower-$M_p$ component to be relatively high and by varying the concentration of the higher-$M_p$ component in the blend, we have demonstrated that the component which dominates the EC thinning dynamics can be switched. The results suggest that, for any given MWD with $\text{PDI}>1$, the dominant mode controlling the EC thinning of polymer solutions will depend on the total concentration of polymer $c$ in the solution. 

The transition into the EC thinning regime depends on the elastic stress dominating over the viscous stress in the self-thinning capillary bridge. The elastic stress that drives this transition is generated by the stretching of polymer during the initial VC thinning regime in which the strain rate increases monotonically with time. We therefore hypothesize that over time progressively shorter polymers in the MWD begin to stretch and contribute to the elastic stress. Depending on the total polymer concentration, a different fraction of the MWD will be required to stretch in order to overcome the viscous stress and induce the onset of EC thinning. Therefore the EC thinning dynamics depend on $c$ since for increasing $c$ a higher molecular weight of polymer with a longer relaxation time is effectively controlling the EC thinning process. 

Based on the assumption of sequential polymer stretching within the VC regime we are able to qualitatively capture our experimental results through a simplified model in which we sum the elastic stress contributions of progressively decreasing modes in the MWD until the accumulated elastic stress exceeds the viscous stress (i.e., until the elastic stress passes the threshold for onset of the EC regime in a capillary thinning experiment). Given the simplicity of the model, and the implicit assumptions therein, we consider the qualitative agreement that it yields with the experiment amounts to compelling support for our hypothesis. This shows that the measured value of $\tau_{EC}$ can depend on the concentration of polymer in solution even in the absence of interpolymer interactions, simply due to the spread of timescales present in the molecular weight distribution. This observation means that defining concentration criteria for ultradiluteness such as $c^*_s$ (Eq.~\ref{c*_s}, \cite{Dinic2020}) may not be necessary in order to understand the concentration dependence of $\tau_{EC}$ in dilute conditions. Instead, at least some degree of concentration dependence should simply be expected in the capillary thinning of polydisperse polymer systems.

As for other concentration criteria, we have shown that the $c_{min}$ criterion (Eq.~\ref{c_min} \cite{Clasen2006}) can be estimated for polydisperse systems and has meaning as the concentration below which an EC regime would not be observed even if the whole of the MWD became fully stretched. However, for polydisperse systems the $c_{low}$ criterion (Eq.~\ref{c_low} \cite{Campo2010}) appears to lose any significance. This criterion defines the minimum concentration of polymer required to induce an EC regime at the onset of stretching, a requirement for the extraction of the `true' relaxation time from the EC thinning regime. However, $c_{low}$ implicity assumes the system of interest is monodisperse and that all of the polymers in the solution will deform at the same extension rate. Considering the $c_{low}$ criterion in context of a MWD, a question arises about which of the many disperse polymers present needs to start stretching in order to induce the EC regime (essentially the question that we address in this paper). Our current results indicate that for any dilute polymer solution yielding an EC regime it should be reasonable to extract a value for $\tau_{EC}$, that should be related to some relaxation timescale of the fluid. Confoundingly, though, the extracted timescale will correspond to a concentration-dependent and, in general unknown, molecular weight fraction of the MWD. 

The current work (and also another recent work involving capillary thinning of solutions of polymers with various flexibilities \cite{Calabrese2024}) reveals the great importance of the initial VC thinning stage prior to the onset of the EC regime. Previous work has shown that significant pre-alignment of stiffer polymeric systems can occur during the VC regime, while the present work shows that significant pre-stretching of more flexible polymers is also possible prior to the onset of elastocapillarity (as also suggested by another recent study \cite{Gaillard2024}). This pre-stretching and pre-alignment determines what remains to be stretched and/or aligned during the EC regime itself and therefore dictates exactly what component of the fluid the extracted value of $\tau_{EC}$ corresponds to. 



As discussed in Ref.~\cite{Calabrese2024}, elastocapillary thinning measurements can provide useful timescales relevant to the observation of elastic effects in extensional flows of complex fluids. However, our current results have clear implications for extensional rheometry techniques based on capillary thinning measurements, and may provide new understanding relevant to the interpretation of some previous results reported in the literature. For instance, the derivation of scaling laws for polymer relaxation times with concentration based on capillary thinning measurements made with highly disperse commercial polymers (e.g., \cite{Dinic2020,Rajesh2022,Soetrisno2023}) is called into question since it is clearly inadequate to relate such measurements to a single molecular weight defined by $M_w$ or $M_p$. Any `concentration dependence' may not be solely due to concentration but may be significantly affected by the MWD. Apart from the concentration dependence of $\tau_{EC}$, it also becomes clear why capillary thinning of very dilute polymer solutions can yield timescales shorter than expected based on estimates made for $M_w$ or $M_p$ (e.g., \cite{Clasen2006,Dinic2020,Soetrisno2023}). More fundamentally, issues related to the fluid's self-selection of the strain rate in the EC thinning regime are brought into sharp clarity. Generally in rheometry it is desirable to impose the deformation rate, and to be able to vary that imposed rate while measuring the stress required to maintain it, an approach that allows a range of fluid timescales to be probed by the rheometerist. For a given polymer sample in the dilute regime, imposition of a controlled deformation rate should result in the same fraction of the MWD of the sample being deformed regardless of concentration. This is obviously not possible in a capillary thinning experiment, which is self-evident since the deformation rate in the EC regime depends on concentration, see Fig.~\ref{SRM_profiles}(c,d). 
In most branches of rheometry it would be implicity understood that for the application of different strain rates, different fluid timescales are being probed in the experiment. At a most basic level, here we are proposing that exactly the same concept applies in the case of capillary thinning-based extensional rheometry techniques. 

\section{Conclusions}

The thinning of fluid filaments under the action of capillary pressure impacts widespread important processes and is utilized for the measurement of characteristic `relaxation times' of complex fluids such as polymer solutions. An important question is why the determination of such characteristic times depends on the polymer concentration even when the fluid is dilute. This is usually explained by invoking intermolecular interactions occuring in fluids that are not `ultradilute', and by `self-concentration' effects as the polymer stretches \cite{Harrison1998,Clasen2006,Dinic2020}. Although relaxation times are undoubtedly affected by interpolymer interactions above some concentration limit (normally considered as being $c>c^*$), our experiments and simple calculations indicate that the polymer molecular weight distribution also requires accounting for and may even fully explain the concentration dependence of elastocapillary timescales observed at concentrations $c<c^*$. 

Our results show how a concentration-dependent portion of a polymer's molecular weight distribution can become stretched during the Newtonian-like thinning stage occurring prior to the onset of elastocapillarity. This then determines which polymers in the distribution will control the rate of thinning during the subsequent elastocapillary stage: for higher polymer concentrations, onset of elastocapillarity occurs at higher molecular weights. Alternatively, by considering the concentration-dependent extension rate achieved during elastocapillary thinning it becomes self-evident that longer fluid time scales (hence higher molecular weights) are being probed by the experiment as the polymer concentration is increased. The observed concentration dependence of elastocapillary timescales is thus a direct consequence of the inability to control the extension rate during capillary-driven self-thinning. This has significant implications for the application of capillary thinning to the aquisition of rheometric data, and for the interpretation of data thus obtained. 

Finally, we reitterate that although we focuss our discussion here on the relevance to extensional rheometry, an improved understanding of the important role of the molecular weight distribution could potentially benefit numerous practical applications that involve the capillary-driven thinning of viscoelastic fluids. We hope that this work should motivate future experimental, numerical, and theoretical studies aimed towards this goal.

\vspace{+0.2in}

We gratefully acknowledge the support of the Okinawa Institute of Science and Technology Graduate University (OIST) with subsidy funding from the Cabinet Office, Government of Japan, and also funding from the Japan Society for the Promotion of Science (JSPS, Grant Nos. 24K07332, 24K17736, and 24K00810). We are indebted to Dr. Isaac Pincus and Prof. Gareth McKinley from MIT for their helpful discussions, suggestions and comments on the manuscript.

\newpage

\bibliographystyle{apsrev4-2}


\begin{thebibliography}{0}%
\makeatletter
\providecommand \@ifxundefined [1]{%
 \@ifx{#1\undefined}
}%
\providecommand \@ifnum [1]{%
 \ifnum #1\expandafter \@firstoftwo
 \else \expandafter \@secondoftwo
 \fi
}%
\providecommand \@ifx [1]{%
 \ifx #1\expandafter \@firstoftwo
 \else \expandafter \@secondoftwo
 \fi
}%
\providecommand \natexlab [1]{#1}%
\providecommand \enquote  [1]{``#1''}%
\providecommand \bibnamefont  [1]{#1}%
\providecommand \bibfnamefont [1]{#1}%
\providecommand \citenamefont [1]{#1}%
\providecommand \href@noop [0]{\@secondoftwo}%
\providecommand \href [0]{\begingroup \@sanitize@url \@href}%
\providecommand \@href[1]{\@@startlink{#1}\@@href}%
\providecommand \@@href[1]{\endgroup#1\@@endlink}%
\providecommand \@sanitize@url [0]{\catcode `\\12\catcode `\$12\catcode
  `\&12\catcode `\#12\catcode `\^12\catcode `\_12\catcode `\%12\relax}%
\providecommand \@@startlink[1]{}%
\providecommand \@@endlink[0]{}%
\providecommand \url  [0]{\begingroup\@sanitize@url \@url }%
\providecommand \@url [1]{\endgroup\@href {#1}{\urlprefix }}%
\providecommand \urlprefix  [0]{URL }%
\providecommand \Eprint [0]{\href }%
\providecommand \doibase [0]{http://dx.doi.org/}%
\providecommand \selectlanguage [0]{\@gobble}%
\providecommand \bibinfo  [0]{\@secondoftwo}%
\providecommand \bibfield  [0]{\@secondoftwo}%
\providecommand \translation [1]{[#1]}%
\providecommand \BibitemOpen [0]{}%
\providecommand \bibitemStop [0]{}%
\providecommand \bibitemNoStop [0]{.\EOS\space}%
\providecommand \EOS [0]{\spacefactor3000\relax}%
\providecommand \BibitemShut  [1]{\csname bibitem#1\endcsname}%
\let\auto@bib@innerbib\@empty
\end{thebibliography}%


\begin{thebibliography}{64}%
\makeatletter
\providecommand \@ifxundefined [1]{%
 \@ifx{#1\undefined}
}%
\providecommand \@ifnum [1]{%
 \ifnum #1\expandafter \@firstoftwo
 \else \expandafter \@secondoftwo
 \fi
}%
\providecommand \@ifx [1]{%
 \ifx #1\expandafter \@firstoftwo
 \else \expandafter \@secondoftwo
 \fi
}%
\providecommand \natexlab [1]{#1}%
\providecommand \enquote  [1]{``#1''}%
\providecommand \bibnamefont  [1]{#1}%
\providecommand \bibfnamefont [1]{#1}%
\providecommand \citenamefont [1]{#1}%
\providecommand \href@noop [0]{\@secondoftwo}%
\providecommand \href [0]{\begingroup \@sanitize@url \@href}%
\providecommand \@href[1]{\@@startlink{#1}\@@href}%
\providecommand \@@href[1]{\endgroup#1\@@endlink}%
\providecommand \@sanitize@url [0]{\catcode `\\12\catcode `\$12\catcode
  `\&12\catcode `\#12\catcode `\^12\catcode `\_12\catcode `\%12\relax}%
\providecommand \@@startlink[1]{}%
\providecommand \@@endlink[0]{}%
\providecommand \url  [0]{\begingroup\@sanitize@url \@url }%
\providecommand \@url [1]{\endgroup\@href {#1}{\urlprefix }}%
\providecommand \urlprefix  [0]{URL }%
\providecommand \Eprint [0]{\href }%
\providecommand \doibase [0]{http://dx.doi.org/}%
\providecommand \selectlanguage [0]{\@gobble}%
\providecommand \bibinfo  [0]{\@secondoftwo}%
\providecommand \bibfield  [0]{\@secondoftwo}%
\providecommand \translation [1]{[#1]}%
\providecommand \BibitemOpen [0]{}%
\providecommand \bibitemStop [0]{}%
\providecommand \bibitemNoStop [0]{.\EOS\space}%
\providecommand \EOS [0]{\spacefactor3000\relax}%
\providecommand \BibitemShut  [1]{\csname bibitem#1\endcsname}%
\let\auto@bib@innerbib\@empty
\bibitem [{\citenamefont {Keshavarz}\ \emph {et~al.}(2016)\citenamefont
  {Keshavarz}, \citenamefont {Houze}, \citenamefont {Moore}, \citenamefont
  {Koerner},\ and\ \citenamefont {McKinley}}]{Keshavarz2016}%
  \BibitemOpen
  \bibfield  {author} {\bibinfo {author} {\bibfnamefont {B.}~\bibnamefont
  {Keshavarz}}, \bibinfo {author} {\bibfnamefont {E.~C.}\ \bibnamefont
  {Houze}}, \bibinfo {author} {\bibfnamefont {J.~R.}\ \bibnamefont {Moore}},
  \bibinfo {author} {\bibfnamefont {M.~R.}\ \bibnamefont {Koerner}}, \ and\
  \bibinfo {author} {\bibfnamefont {G.~H.}\ \bibnamefont {McKinley}},\
  }\bibfield  {title} {\emph {\enquote {\bibinfo {title} {Ligament mediated
  fragmentation of viscoelastic liquids},}\ }}\href@noop {} {\bibfield
  {journal} {\bibinfo  {journal} {Phys. Rev. Lett.}\ }\textbf {\bibinfo
  {volume} {117}},\ \bibinfo {pages} {154502} (\bibinfo {year}
  {2016})}\BibitemShut {NoStop}%
\bibitem [{\citenamefont {Keshavarz}\ \emph {et~al.}(2020)\citenamefont
  {Keshavarz}, \citenamefont {Houze}, \citenamefont {Moore}, \citenamefont
  {Koerner},\ and\ \citenamefont {McKinley}}]{Keshavarz2020}%
  \BibitemOpen
  \bibfield  {author} {\bibinfo {author} {\bibfnamefont {B.}~\bibnamefont
  {Keshavarz}}, \bibinfo {author} {\bibfnamefont {E.~C.}\ \bibnamefont
  {Houze}}, \bibinfo {author} {\bibfnamefont {J.~R.}\ \bibnamefont {Moore}},
  \bibinfo {author} {\bibfnamefont {M.~R.}\ \bibnamefont {Koerner}}, \ and\
  \bibinfo {author} {\bibfnamefont {G.~H.}\ \bibnamefont {McKinley}},\
  }\bibfield  {title} {\emph {\enquote {\bibinfo {title} {Rotary atomization of
  {N}ewtonian and viscoelastic liquids},}\ }}\href {\doibase
  10.1103/PhysRevFluids.5.033601} {\bibfield  {journal} {\bibinfo  {journal}
  {Phys. Rev. Fluids}\ }\textbf {\bibinfo {volume} {5}},\ \bibinfo {pages}
  {033601} (\bibinfo {year} {2020})}\BibitemShut {NoStop}%
\bibitem [{\citenamefont {Amarouchene}\ \emph {et~al.}(2001)\citenamefont
  {Amarouchene}, \citenamefont {Bonn}, \citenamefont {Meunier},\ and\
  \citenamefont {Kellay}}]{Amarouchene2001}%
  \BibitemOpen
  \bibfield  {author} {\bibinfo {author} {\bibfnamefont {Y.}~\bibnamefont
  {Amarouchene}}, \bibinfo {author} {\bibfnamefont {D.}~\bibnamefont {Bonn}},
  \bibinfo {author} {\bibfnamefont {J.}~\bibnamefont {Meunier}}, \ and\
  \bibinfo {author} {\bibfnamefont {H.}~\bibnamefont {Kellay}},\ }\bibfield
  {title} {\emph {\enquote {\bibinfo {title} {Inhibition of the finite-time
  singularity during droplet fission of a polymeric fluid},}\ }}\href {\doibase
  10.1103/PhysRevLett.86.3558} {\bibfield  {journal} {\bibinfo  {journal}
  {Phys. Rev. Lett.}\ }\textbf {\bibinfo {volume} {86}},\ \bibinfo {pages}
  {3558} (\bibinfo {year} {2001})}\BibitemShut {NoStop}%
\bibitem [{\citenamefont {Ingremeau}\ and\ \citenamefont
  {Kellay}(2013)}]{Ingremeau2013}%
  \BibitemOpen
  \bibfield  {author} {\bibinfo {author} {\bibfnamefont {F.}~\bibnamefont
  {Ingremeau}}\ and\ \bibinfo {author} {\bibfnamefont {H.}~\bibnamefont
  {Kellay}},\ }\bibfield  {title} {\emph {\enquote {\bibinfo {title}
  {Stretching polymers in droplet-pinch-off experiments},}\ }}\href@noop {}
  {\bibfield  {journal} {\bibinfo  {journal} {Phys. Rev. X}\ }\textbf {\bibinfo
  {volume} {3}},\ \bibinfo {pages} {041002} (\bibinfo {year}
  {2013})}\BibitemShut {NoStop}%
\bibitem [{\citenamefont {Xue}\ \emph {et~al.}(2019)\citenamefont {Xue},
  \citenamefont {Wu}, \citenamefont {Dai},\ and\ \citenamefont
  {Xia}}]{Xue2019}%
  \BibitemOpen
  \bibfield  {author} {\bibinfo {author} {\bibfnamefont {J.}~\bibnamefont
  {Xue}}, \bibinfo {author} {\bibfnamefont {T.}~\bibnamefont {Wu}}, \bibinfo
  {author} {\bibfnamefont {Y.}~\bibnamefont {Dai}}, \ and\ \bibinfo {author}
  {\bibfnamefont {Y.}~\bibnamefont {Xia}},\ }\bibfield  {title} {\emph
  {\enquote {\bibinfo {title} {Electrospinning and electrospun nanofibers:
  {M}ethods, materials, and applications},}\ }}\href {\doibase
  10.1021/acs.chemrev.8b00593} {\bibfield  {journal} {\bibinfo  {journal}
  {Chem. Rev.}\ }\textbf {\bibinfo {volume} {119}},\ \bibinfo {pages} {5298}
  (\bibinfo {year} {2019})}\BibitemShut {NoStop}%
\bibitem [{\citenamefont {Xu}\ \emph {et~al.}(2022)\citenamefont {Xu},
  \citenamefont {Wu}, \citenamefont {Ye}, \citenamefont {Chen}, \citenamefont
  {Liu},\ and\ \citenamefont {Bai}}]{Xu2022}%
  \BibitemOpen
  \bibfield  {author} {\bibinfo {author} {\bibfnamefont {Z.}~\bibnamefont
  {Xu}}, \bibinfo {author} {\bibfnamefont {M.}~\bibnamefont {Wu}}, \bibinfo
  {author} {\bibfnamefont {Q.}~\bibnamefont {Ye}}, \bibinfo {author}
  {\bibfnamefont {D.}~\bibnamefont {Chen}}, \bibinfo {author} {\bibfnamefont
  {K.}~\bibnamefont {Liu}}, \ and\ \bibinfo {author} {\bibfnamefont
  {H.}~\bibnamefont {Bai}},\ }\bibfield  {title} {\emph {\enquote {\bibinfo
  {title} {Spinning from nature: {E}ngineered preparation and application of
  high-performance bio-based fibers},}\ }}\href {\doibase
  https://doi.org/10.1016/j.eng.2021.06.030} {\bibfield  {journal} {\bibinfo
  {journal} {Engineering - London}\ }\textbf {\bibinfo {volume} {14}},\
  \bibinfo {pages} {100} (\bibinfo {year} {2022})}\BibitemShut {NoStop}%
\bibitem [{\citenamefont {Hutchings}\ and\ \citenamefont
  {Martin}(2013)}]{Hutchings2013}%
  \BibitemOpen
  \bibinfo {editor} {\bibfnamefont {I.~M.}\ \bibnamefont {Hutchings}}\ and\
  \bibinfo {editor} {\bibfnamefont {G.~D.}\ \bibnamefont {Martin}},\ eds.,\
  \href@noop {} {\emph {\bibinfo {title} {Inkjet Technology for Digital
  Fabrication}}}\ (\bibinfo  {publisher} {Wiley},\ \bibinfo {address} {UK},\
  \bibinfo {year} {2013})\BibitemShut {NoStop}%
\bibitem [{\citenamefont {McIlroy}\ \emph {et~al.}(2013)\citenamefont
  {McIlroy}, \citenamefont {Harlen},\ and\ \citenamefont
  {Morrison}}]{McIlroy2013}%
  \BibitemOpen
  \bibfield  {author} {\bibinfo {author} {\bibfnamefont {C.}~\bibnamefont
  {McIlroy}}, \bibinfo {author} {\bibfnamefont {O.~G.}\ \bibnamefont {Harlen}},
  \ and\ \bibinfo {author} {\bibfnamefont {N.~F.}\ \bibnamefont {Morrison}},\
  }\bibfield  {title} {\emph {\enquote {\bibinfo {title} {Modelling the jetting
  of dilute polymer solutions in drop-on-demand inkjet printing},}\ }}\href
  {\doibase https://doi.org/10.1016/j.jnnfm.2013.05.007} {\bibfield  {journal}
  {\bibinfo  {journal} {J. Non-Newt. Fluid Mech.}\ }\textbf {\bibinfo {volume}
  {201}},\ \bibinfo {pages} {17} (\bibinfo {year} {2013})}\BibitemShut
  {NoStop}%
\bibitem [{\citenamefont {Lohse}(2022)}]{Lohse2022}%
  \BibitemOpen
  \bibfield  {author} {\bibinfo {author} {\bibfnamefont {D.}~\bibnamefont
  {Lohse}},\ }\bibfield  {title} {\emph {\enquote {\bibinfo {title}
  {Fundamental fluid dynamics challenges in inkjet printing},}\ }}\href
  {\doibase 10.1146/annurev-fluid-022321-114001} {\bibfield  {journal}
  {\bibinfo  {journal} {Annu. Rev. Fluid Mech.}\ }\textbf {\bibinfo {volume}
  {54}},\ \bibinfo {pages} {349} (\bibinfo {year} {2022})}\BibitemShut
  {NoStop}%
\bibitem [{\citenamefont {Wei}\ \emph {et~al.}(2015)\citenamefont {Wei},
  \citenamefont {Li}, \citenamefont {David}, \citenamefont {Jones},
  \citenamefont {Sarohia}, \citenamefont {Schmitigal},\ and\ \citenamefont
  {Kornfield}}]{Wei2015}%
  \BibitemOpen
  \bibfield  {author} {\bibinfo {author} {\bibfnamefont {M.-N.}\ \bibnamefont
  {Wei}}, \bibinfo {author} {\bibfnamefont {B.}~\bibnamefont {Li}}, \bibinfo
  {author} {\bibfnamefont {R.~L.~A.}\ \bibnamefont {David}}, \bibinfo {author}
  {\bibfnamefont {S.~C.}\ \bibnamefont {Jones}}, \bibinfo {author}
  {\bibfnamefont {V.}~\bibnamefont {Sarohia}}, \bibinfo {author} {\bibfnamefont
  {J.~A.}\ \bibnamefont {Schmitigal}}, \ and\ \bibinfo {author} {\bibfnamefont
  {J.~A.}\ \bibnamefont {Kornfield}},\ }\bibfield  {title} {\emph {\enquote
  {\bibinfo {title} {Megasupramolecules for safer, cleaner fuel by end
  association of long telechelic polymers},}\ }}\href@noop {} {\bibfield
  {journal} {\bibinfo  {journal} {Science}\ }\textbf {\bibinfo {volume}
  {350}},\ \bibinfo {pages} {72} (\bibinfo {year} {2015})}\BibitemShut
  {NoStop}%
\bibitem [{\citenamefont {Scharfman}\ \emph {et~al.}(2016)\citenamefont
  {Scharfman}, \citenamefont {Techet}, \citenamefont {Bush},\ and\
  \citenamefont {Bourouiba}}]{Scharfman2016}%
  \BibitemOpen
  \bibfield  {author} {\bibinfo {author} {\bibfnamefont {B.~E.}\ \bibnamefont
  {Scharfman}}, \bibinfo {author} {\bibfnamefont {A.~H.}\ \bibnamefont
  {Techet}}, \bibinfo {author} {\bibfnamefont {J.~W.~M.}\ \bibnamefont {Bush}},
  \ and\ \bibinfo {author} {\bibfnamefont {L.}~\bibnamefont {Bourouiba}},\
  }\bibfield  {title} {\emph {\enquote {\bibinfo {title} {Visualization of
  sneeze ejecta: steps of fluid fragmentation leading to respiratory
  droplets},}\ }}\href {\doibase 10.1007/s00348-015-2078-4} {\bibfield
  {journal} {\bibinfo  {journal} {Exp. Fluids}\ }\textbf {\bibinfo {volume}
  {57}},\ \bibinfo {pages} {24} (\bibinfo {year} {2016})}\BibitemShut {NoStop}%
\bibitem [{\citenamefont {Abkarian}\ and\ \citenamefont
  {Stone}(2020)}]{Abkarian2020}%
  \BibitemOpen
  \bibfield  {author} {\bibinfo {author} {\bibfnamefont {M.}~\bibnamefont
  {Abkarian}}\ and\ \bibinfo {author} {\bibfnamefont {H.~A.}\ \bibnamefont
  {Stone}},\ }\bibfield  {title} {\emph {\enquote {\bibinfo {title} {Stretching
  and break-up of saliva filaments during speech: {A} route for pathogen
  aerosolization and its potential mitigation},}\ }}\href {\doibase
  10.1103/PhysRevFluids.5.102301} {\bibfield  {journal} {\bibinfo  {journal}
  {Phys. Rev. Fluids}\ }\textbf {\bibinfo {volume} {5}},\ \bibinfo {pages}
  {102301} (\bibinfo {year} {2020})}\BibitemShut {NoStop}%
\bibitem [{\citenamefont {Bourouiba}(2021)}]{Bourouiba2021}%
  \BibitemOpen
  \bibfield  {author} {\bibinfo {author} {\bibfnamefont {L.}~\bibnamefont
  {Bourouiba}},\ }\bibfield  {title} {\emph {\enquote {\bibinfo {title} {The
  fluid dynamics of disease transmission},}\ }}\href {\doibase
  10.1146/annurev-fluid-060220-113712} {\bibfield  {journal} {\bibinfo
  {journal} {Annu. Rev. Fluid Mech.}\ }\textbf {\bibinfo {volume} {53}},\
  \bibinfo {pages} {473} (\bibinfo {year} {2021})}\BibitemShut {NoStop}%
\bibitem [{\citenamefont {Papageorgiou}(1995)}]{Papageorgiou1995}%
  \BibitemOpen
  \bibfield  {author} {\bibinfo {author} {\bibfnamefont {D.~T.}\ \bibnamefont
  {Papageorgiou}},\ }\bibfield  {title} {\emph {\enquote {\bibinfo {title} {On
  the breakup of viscous liquid threads},}\ }}\href@noop {} {\bibfield
  {journal} {\bibinfo  {journal} {Phys. Fluids}\ }\textbf {\bibinfo {volume}
  {7}},\ \bibinfo {pages} {1529} (\bibinfo {year} {1995})}\BibitemShut
  {NoStop}%
\bibitem [{\citenamefont {McKinley}\ and\ \citenamefont
  {Tripathi}(2000)}]{McKinley2000}%
  \BibitemOpen
  \bibfield  {author} {\bibinfo {author} {\bibfnamefont {G.~H.}\ \bibnamefont
  {McKinley}}\ and\ \bibinfo {author} {\bibfnamefont {A.}~\bibnamefont
  {Tripathi}},\ }\bibfield  {title} {\emph {\enquote {\bibinfo {title} {How to
  extract the {N}ewtonian viscosity from capillary breakup measurements in a
  filament rheometer},}\ }}\href {\doibase 10.1122/1.551105} {\bibfield
  {journal} {\bibinfo  {journal} {J. Rheol.}\ }\textbf {\bibinfo {volume}
  {44}},\ \bibinfo {pages} {653} (\bibinfo {year} {2000})}\BibitemShut
  {NoStop}%
\bibitem [{\citenamefont {Entov}\ and\ \citenamefont
  {Hinch}(1997)}]{Entov1997}%
  \BibitemOpen
  \bibfield  {author} {\bibinfo {author} {\bibfnamefont {V.~M.}\ \bibnamefont
  {Entov}}\ and\ \bibinfo {author} {\bibfnamefont {E.~J.}\ \bibnamefont
  {Hinch}},\ }\bibfield  {title} {\emph {\enquote {\bibinfo {title} {Effect of
  a spectrum of relaxation times on the capillary thinning of a filament of
  elastic liquid},}\ }}\href {\doibase 10.1016/S0377-0257(97)00022-0}
  {\bibfield  {journal} {\bibinfo  {journal} {J. Non-Newt. Fluid Mech.}\
  }\textbf {\bibinfo {volume} {72}},\ \bibinfo {pages} {31} (\bibinfo {year}
  {1997})}\BibitemShut {NoStop}%
\bibitem [{\citenamefont {Anna}\ and\ \citenamefont
  {McKinley}(2001)}]{Anna2001b}%
  \BibitemOpen
  \bibfield  {author} {\bibinfo {author} {\bibfnamefont {S.~L.}\ \bibnamefont
  {Anna}}\ and\ \bibinfo {author} {\bibfnamefont {G.~H.}\ \bibnamefont
  {McKinley}},\ }\bibfield  {title} {\emph {\enquote {\bibinfo {title}
  {Elasto-capillary thinning and breakup of model elastic liquids},}\ }}\href
  {\doibase 10.1122/1.1332389} {\bibfield  {journal} {\bibinfo  {journal} {J.
  Rheol.}\ }\textbf {\bibinfo {volume} {45}},\ \bibinfo {pages} {115} (\bibinfo
  {year} {2001})}\BibitemShut {NoStop}%
\bibitem [{\citenamefont {Clasen}\ \emph {et~al.}(2006)\citenamefont {Clasen},
  \citenamefont {Plog}, \citenamefont {Kulicke}, \citenamefont {Owens},
  \citenamefont {Macosko}, \citenamefont {Scriven}, \citenamefont {Verani},\
  and\ \citenamefont {McKinley}}]{Clasen2006}%
  \BibitemOpen
  \bibfield  {author} {\bibinfo {author} {\bibfnamefont {C.}~\bibnamefont
  {Clasen}}, \bibinfo {author} {\bibfnamefont {J.~P.}\ \bibnamefont {Plog}},
  \bibinfo {author} {\bibfnamefont {W.-M.}\ \bibnamefont {Kulicke}}, \bibinfo
  {author} {\bibfnamefont {M.}~\bibnamefont {Owens}}, \bibinfo {author}
  {\bibfnamefont {C.}~\bibnamefont {Macosko}}, \bibinfo {author} {\bibfnamefont
  {L.~E.}\ \bibnamefont {Scriven}}, \bibinfo {author} {\bibfnamefont
  {M.}~\bibnamefont {Verani}}, \ and\ \bibinfo {author} {\bibfnamefont {G.~H.}\
  \bibnamefont {McKinley}},\ }\bibfield  {title} {\emph {\enquote {\bibinfo
  {title} {How dilute are dilute solutions in extensional flows?}}\ }}\href
  {\doibase 10.1122/1.2357595} {\bibfield  {journal} {\bibinfo  {journal} {J.
  Rheol.}\ }\textbf {\bibinfo {volume} {50}},\ \bibinfo {pages} {849} (\bibinfo
  {year} {2006})}\BibitemShut {NoStop}%
\bibitem [{\citenamefont {Campo-Dea\~no}\ and\ \citenamefont
  {Clasen}(2010)}]{Campo2010}%
  \BibitemOpen
  \bibfield  {author} {\bibinfo {author} {\bibfnamefont {L.}~\bibnamefont
  {Campo-Dea\~no}}\ and\ \bibinfo {author} {\bibfnamefont {C.}~\bibnamefont
  {Clasen}},\ }\bibfield  {title} {\emph {\enquote {\bibinfo {title} {The slow
  retraction method ({SRM}) for the determination of ultra-short relaxation
  times in capillary breakup extensional rheometry experiments},}\ }}\href
  {\doibase 10.1016/j.jnnfm.2010.09.007} {\bibfield  {journal} {\bibinfo
  {journal} {J. Non-Newt. Fluid Mech.}\ }\textbf {\bibinfo {volume} {165}},\
  \bibinfo {pages} {1688} (\bibinfo {year} {2010})}\BibitemShut {NoStop}%
\bibitem [{\citenamefont {Nelson}\ \emph {et~al.}(2011)\citenamefont {Nelson},
  \citenamefont {Kavehpour},\ and\ \citenamefont {Kim}}]{Nelson2011}%
  \BibitemOpen
  \bibfield  {author} {\bibinfo {author} {\bibfnamefont {W.~C.}\ \bibnamefont
  {Nelson}}, \bibinfo {author} {\bibfnamefont {H.~P.}\ \bibnamefont
  {Kavehpour}}, \ and\ \bibinfo {author} {\bibfnamefont {C.~J.}\ \bibnamefont
  {Kim}},\ }\bibfield  {title} {\emph {\enquote {\bibinfo {title} {A miniature
  capillary breakup extensional rheometer by electrostatically assisted
  generation of liquid filaments},}\ }}\href@noop {} {\bibfield  {journal}
  {\bibinfo  {journal} {Lab Chip}\ }\textbf {\bibinfo {volume} {11}},\ \bibinfo
  {pages} {2424} (\bibinfo {year} {2011})}\BibitemShut {NoStop}%
\bibitem [{\citenamefont {Bhattacharjee}\ \emph {et~al.}(2011)\citenamefont
  {Bhattacharjee}, \citenamefont {McDonnell}, \citenamefont {Prabhakar},
  \citenamefont {Yeo},\ and\ \citenamefont {Friend}}]{Bhattacharjee2011}%
  \BibitemOpen
  \bibfield  {author} {\bibinfo {author} {\bibfnamefont {P.~K.}\ \bibnamefont
  {Bhattacharjee}}, \bibinfo {author} {\bibfnamefont {A.~G.}\ \bibnamefont
  {McDonnell}}, \bibinfo {author} {\bibfnamefont {R.}~\bibnamefont
  {Prabhakar}}, \bibinfo {author} {\bibfnamefont {L.~Y.}\ \bibnamefont {Yeo}},
  \ and\ \bibinfo {author} {\bibfnamefont {J.~R.}\ \bibnamefont {Friend}},\
  }\bibfield  {title} {\emph {\enquote {\bibinfo {title} {Extensional flow of
  low-viscosity fluids in capillary bridges formed by pulsed surface acoustic
  wave jetting},}\ }}\href {\doibase 10.1088/1367-2630/13/2/023005} {\bibfield
  {journal} {\bibinfo  {journal} {New J. Phys.}\ }\textbf {\bibinfo {volume}
  {13}},\ \bibinfo {pages} {023005} (\bibinfo {year} {2011})}\BibitemShut
  {NoStop}%
\bibitem [{\citenamefont {Dinic}\ \emph {et~al.}(2015)\citenamefont {Dinic},
  \citenamefont {Zhang}, \citenamefont {Jimenez},\ and\ \citenamefont
  {Sharma}}]{Dinic2015}%
  \BibitemOpen
  \bibfield  {author} {\bibinfo {author} {\bibfnamefont {J.}~\bibnamefont
  {Dinic}}, \bibinfo {author} {\bibfnamefont {Y.}~\bibnamefont {Zhang}},
  \bibinfo {author} {\bibfnamefont {L.~N.}\ \bibnamefont {Jimenez}}, \ and\
  \bibinfo {author} {\bibfnamefont {V.}~\bibnamefont {Sharma}},\ }\bibfield
  {title} {\emph {\enquote {\bibinfo {title} {Extensional relaxation times of
  dilute, aqueous polymer solutions},}\ }}\href {\doibase
  10.1021/acsmacrolett.5b00393} {\bibfield  {journal} {\bibinfo  {journal} {ACS
  Macro Lett.}\ }\textbf {\bibinfo {volume} {4}},\ \bibinfo {pages} {804}
  (\bibinfo {year} {2015})}\BibitemShut {NoStop}%
\bibitem [{\citenamefont {Keshavarz}\ \emph {et~al.}(2015)\citenamefont
  {Keshavarz}, \citenamefont {Sharma}, \citenamefont {Houze}, \citenamefont
  {Koerner}, \citenamefont {Moore}, \citenamefont {Cotts}, \citenamefont
  {Threlfall-Holmes},\ and\ \citenamefont {McKinley}}]{Keshavarz2015}%
  \BibitemOpen
  \bibfield  {author} {\bibinfo {author} {\bibfnamefont {B.}~\bibnamefont
  {Keshavarz}}, \bibinfo {author} {\bibfnamefont {V.}~\bibnamefont {Sharma}},
  \bibinfo {author} {\bibfnamefont {E.~C.}\ \bibnamefont {Houze}}, \bibinfo
  {author} {\bibfnamefont {M.~R.}\ \bibnamefont {Koerner}}, \bibinfo {author}
  {\bibfnamefont {J.~R.}\ \bibnamefont {Moore}}, \bibinfo {author}
  {\bibfnamefont {P.~M.}\ \bibnamefont {Cotts}}, \bibinfo {author}
  {\bibfnamefont {P.}~\bibnamefont {Threlfall-Holmes}}, \ and\ \bibinfo
  {author} {\bibfnamefont {G.~H.}\ \bibnamefont {McKinley}},\ }\bibfield
  {title} {\emph {\enquote {\bibinfo {title} {Studying the effects of
  elongational properties on atomization of weakly viscoelastic solutions using
  {R}ayleigh {O}hnesorge {J}etting {E}xtensional {R}heometry ({ROJER})},}\
  }}\href {\doibase 10.1016/j.jnnfm.2014.11.004} {\bibfield  {journal}
  {\bibinfo  {journal} {J. Non-Newt. Fluid Mech.}\ }\textbf {\bibinfo {volume}
  {222}},\ \bibinfo {pages} {171} (\bibinfo {year} {2015})}\BibitemShut
  {NoStop}%
\bibitem [{\citenamefont {Sharma}\ \emph {et~al.}(2015)\citenamefont {Sharma},
  \citenamefont {Haward}, \citenamefont {Serdy}, \citenamefont {Keshavarz},
  \citenamefont {Soderlund}, \citenamefont {Threlfall-Holmes},\ and\
  \citenamefont {McKinley}}]{Sharma2015}%
  \BibitemOpen
  \bibfield  {author} {\bibinfo {author} {\bibfnamefont {V.}~\bibnamefont
  {Sharma}}, \bibinfo {author} {\bibfnamefont {S.~J.}\ \bibnamefont {Haward}},
  \bibinfo {author} {\bibfnamefont {J.}~\bibnamefont {Serdy}}, \bibinfo
  {author} {\bibfnamefont {B.}~\bibnamefont {Keshavarz}}, \bibinfo {author}
  {\bibfnamefont {A.}~\bibnamefont {Soderlund}}, \bibinfo {author}
  {\bibfnamefont {P.}~\bibnamefont {Threlfall-Holmes}}, \ and\ \bibinfo
  {author} {\bibfnamefont {G.~H.}\ \bibnamefont {McKinley}},\ }\bibfield
  {title} {\emph {\enquote {\bibinfo {title} {The rheology of aqueous solutions
  of ethyl hydroxy-ethyl cellulose ({EHEC}) and its hydrophobically modified
  analogue (hm{EHEC}): {E}xtensional flow response in capillary break-up,
  jetting ({ROJER}) and in a cross-slot extensional rheometer},}\ }}\href
  {\doibase 10.1039/c4sm01661k} {\bibfield  {journal} {\bibinfo  {journal}
  {Soft Matter}\ }\textbf {\bibinfo {volume} {11}},\ \bibinfo {pages} {3251}
  (\bibinfo {year} {2015})}\BibitemShut {NoStop}%
\bibitem [{\citenamefont {Mathues}\ \emph {et~al.}(2018)\citenamefont
  {Mathues}, \citenamefont {Formenti}, \citenamefont {McIlroy}, \citenamefont
  {Harlen},\ and\ \citenamefont {Clasen}}]{Mathues2018}%
  \BibitemOpen
  \bibfield  {author} {\bibinfo {author} {\bibfnamefont {W.}~\bibnamefont
  {Mathues}}, \bibinfo {author} {\bibfnamefont {S.}~\bibnamefont {Formenti}},
  \bibinfo {author} {\bibfnamefont {C.}~\bibnamefont {McIlroy}}, \bibinfo
  {author} {\bibfnamefont {O.~G.}\ \bibnamefont {Harlen}}, \ and\ \bibinfo
  {author} {\bibfnamefont {C.}~\bibnamefont {Clasen}},\ }\bibfield  {title}
  {\emph {\enquote {\bibinfo {title} {{CaBER vs ROJER}—{D}ifferent time
  scales for the thinning of a weakly elastic jet},}\ }}\href {\doibase
  10.1122/1.5021834} {\bibfield  {journal} {\bibinfo  {journal} {J. Rheol.}\
  }\textbf {\bibinfo {volume} {62}},\ \bibinfo {pages} {1135} (\bibinfo {year}
  {2018})}\BibitemShut {NoStop}%
\bibitem [{\citenamefont {Rajesh}\ \emph {et~al.}(2022)\citenamefont {Rajesh},
  \citenamefont {Thi\'{e}venaz},\ and\ \citenamefont {Sauret}}]{Rajesh2022}%
  \BibitemOpen
  \bibfield  {author} {\bibinfo {author} {\bibfnamefont {S.}~\bibnamefont
  {Rajesh}}, \bibinfo {author} {\bibfnamefont {V.}~\bibnamefont
  {Thi\'{e}venaz}}, \ and\ \bibinfo {author} {\bibfnamefont {A.}~\bibnamefont
  {Sauret}},\ }\bibfield  {title} {\emph {\enquote {\bibinfo {title}
  {Transition to the viscoelastic regime in the thinning of polymer
  solutions},}\ }}\href {\doibase 10.1039/D2SM00202G} {\bibfield  {journal}
  {\bibinfo  {journal} {Soft Matter}\ }\textbf {\bibinfo {volume} {18}},\
  \bibinfo {pages} {3147} (\bibinfo {year} {2022})}\BibitemShut {NoStop}%
\bibitem [{\citenamefont {Gaillard}\ \emph {et~al.}(2023)\citenamefont
  {Gaillard}, \citenamefont {Herrada}, \citenamefont {Deblais}, \citenamefont
  {Eggers},\ and\ \citenamefont {Bonn}}]{Gaillard2023}%
  \BibitemOpen
  \bibfield  {author} {\bibinfo {author} {\bibfnamefont {A.}~\bibnamefont
  {Gaillard}}, \bibinfo {author} {\bibfnamefont {M.~A.}\ \bibnamefont
  {Herrada}}, \bibinfo {author} {\bibfnamefont {A.}~\bibnamefont {Deblais}},
  \bibinfo {author} {\bibfnamefont {J.}~\bibnamefont {Eggers}}, \ and\ \bibinfo
  {author} {\bibfnamefont {D.}~\bibnamefont {Bonn}},\ }\href@noop {} {\enquote
  {\bibinfo {title} {Beware of {CaBER}: {F}ilament thinning rheometry doesn't
  give `the' relaxation time of polymer solutions},}\ } (\bibinfo {year}
  {2023}),\ \Eprint {http://arxiv.org/abs/2309.08440v3} {arXiv:2309.08440v3
  [cond-mat.soft]} \BibitemShut {NoStop}%
\bibitem [{\citenamefont {McKinley}(2005)}]{McKinley2005}%
  \BibitemOpen
  \bibfield  {author} {\bibinfo {author} {\bibfnamefont {G.~H.}\ \bibnamefont
  {McKinley}},\ }in\ \href@noop {} {\emph {\bibinfo {booktitle} {Annual
  Rheology Reviews}}},\ Vol.~\bibinfo {volume} {3},\ \bibinfo {editor} {edited
  by\ \bibinfo {editor} {\bibfnamefont {D.~M.}\ \bibnamefont {Binding}}\ and\
  \bibinfo {editor} {\bibfnamefont {K.}~\bibnamefont {Walters}}}\ (\bibinfo
  {publisher} {British Society of Rheology},\ \bibinfo {address}
  {Aberystwyth},\ \bibinfo {year} {2005})\ pp.\ \bibinfo {pages}
  {1--48}\BibitemShut {NoStop}%
\bibitem [{\citenamefont {Tirtaatmadja}\ \emph {et~al.}(2006)\citenamefont
  {Tirtaatmadja}, \citenamefont {McKinley},\ and\ \citenamefont
  {Cooper-White}}]{Tirtaatmadja2006}%
  \BibitemOpen
  \bibfield  {author} {\bibinfo {author} {\bibfnamefont {V.}~\bibnamefont
  {Tirtaatmadja}}, \bibinfo {author} {\bibfnamefont {G.~H.}\ \bibnamefont
  {McKinley}}, \ and\ \bibinfo {author} {\bibfnamefont {J.~J.}\ \bibnamefont
  {Cooper-White}},\ }\bibfield  {title} {\emph {\enquote {\bibinfo {title}
  {Drop formation and breakup of low viscosity elastic fluids: {E}ffects of
  molecular weight and concentration},}\ }}\href {\doibase 10.1063/1.2190469}
  {\bibfield  {journal} {\bibinfo  {journal} {Phys. Fluids}\ }\textbf {\bibinfo
  {volume} {18}},\ \bibinfo {pages} {043101} (\bibinfo {year}
  {2006})}\BibitemShut {NoStop}%
\bibitem [{\citenamefont {Dinic}\ and\ \citenamefont
  {Sharma}(2019)}]{Dinic2019}%
  \BibitemOpen
  \bibfield  {author} {\bibinfo {author} {\bibfnamefont {J.}~\bibnamefont
  {Dinic}}\ and\ \bibinfo {author} {\bibfnamefont {V.}~\bibnamefont {Sharma}},\
  }\bibfield  {title} {\emph {\enquote {\bibinfo {title} {Macromolecular
  relaxation, strain, and extensibility determine elastocapillary thinning and
  extensional viscosity of polymer solutions},}\ }}\href {\doibase
  10.1073/pnas.1820277116} {\bibfield  {journal} {\bibinfo  {journal} {Proc.
  Natl. Acad. Sci. USA}\ }\textbf {\bibinfo {volume} {116}},\ \bibinfo {pages}
  {8766} (\bibinfo {year} {2019})}\BibitemShut {NoStop}%
\bibitem [{\citenamefont {Colby}(2023)}]{Colby2023}%
  \BibitemOpen
  \bibfield  {author} {\bibinfo {author} {\bibfnamefont {R.~H.}\ \bibnamefont
  {Colby}},\ }\bibfield  {title} {\emph {\enquote {\bibinfo {title} {Fiber
  spinning from polymer solutions},}\ }}\href {\doibase 10.1122/8.0000726}
  {\bibfield  {journal} {\bibinfo  {journal} {J. Rheol.}\ }\textbf {\bibinfo
  {volume} {67}},\ \bibinfo {pages} {1251} (\bibinfo {year}
  {2023})}\BibitemShut {NoStop}%
\bibitem [{\citenamefont {Soetrisno}\ \emph {et~al.}(2023)\citenamefont
  {Soetrisno}, \citenamefont {Martínez~Narváez}, \citenamefont {Sharma},\
  and\ \citenamefont {Conrad}}]{Soetrisno2023}%
  \BibitemOpen
  \bibfield  {author} {\bibinfo {author} {\bibfnamefont {D.~D.}\ \bibnamefont
  {Soetrisno}}, \bibinfo {author} {\bibfnamefont {C.~D.~V.}\ \bibnamefont
  {Martínez~Narváez}}, \bibinfo {author} {\bibfnamefont {V.}~\bibnamefont
  {Sharma}}, \ and\ \bibinfo {author} {\bibfnamefont {J.~C.}\ \bibnamefont
  {Conrad}},\ }\bibfield  {title} {\emph {\enquote {\bibinfo {title}
  {Concentration regimes for extensional relaxation times of unentangled
  polymer solutions},}\ }}\href {\doibase 10.1021/acs.macromol.3c00097}
  {\bibfield  {journal} {\bibinfo  {journal} {Macromolecules}\ }\textbf
  {\bibinfo {volume} {56}},\ \bibinfo {pages} {4919} (\bibinfo {year}
  {2023})}\BibitemShut {NoStop}%
\bibitem [{\citenamefont {Harrison}\ \emph {et~al.}(1998)\citenamefont
  {Harrison}, \citenamefont {Remmelgas},\ and\ \citenamefont
  {Leal}}]{Harrison1998}%
  \BibitemOpen
  \bibfield  {author} {\bibinfo {author} {\bibfnamefont {G.~M.}\ \bibnamefont
  {Harrison}}, \bibinfo {author} {\bibfnamefont {J.}~\bibnamefont {Remmelgas}},
  \ and\ \bibinfo {author} {\bibfnamefont {L.~G.}\ \bibnamefont {Leal}},\
  }\bibfield  {title} {\emph {\enquote {\bibinfo {title} {The dynamics of
  ultradilute polymer solutions in transient flow: Comparison of dumbbell-based
  theory and experiment},}\ }}\href {\doibase 10.1122/1.550924} {\bibfield
  {journal} {\bibinfo  {journal} {J. Rheol.}\ }\textbf {\bibinfo {volume}
  {42}},\ \bibinfo {pages} {1039} (\bibinfo {year} {1998})}\BibitemShut
  {NoStop}%
\bibitem [{\citenamefont {Dinic}\ and\ \citenamefont
  {Sharma}(2020)}]{Dinic2020}%
  \BibitemOpen
  \bibfield  {author} {\bibinfo {author} {\bibfnamefont {J.}~\bibnamefont
  {Dinic}}\ and\ \bibinfo {author} {\bibfnamefont {V.}~\bibnamefont {Sharma}},\
  }\bibfield  {title} {\emph {\enquote {\bibinfo {title} {Flexibility,
  extensibility, and ratio of {K}uhn length to packing length govern the
  pinching dynamics, coil-stretch transition, and rheology of polymer
  solutions},}\ }}\href {\doibase 10.1021/acs.macromol.0c00076} {\bibfield
  {journal} {\bibinfo  {journal} {Macromolecules}\ }\textbf {\bibinfo {volume}
  {53}},\ \bibinfo {pages} {4821} (\bibinfo {year} {2020})}\BibitemShut
  {NoStop}%
\bibitem [{\citenamefont {Middleman}(1967)}]{Middleman1967}%
  \BibitemOpen
  \bibfield  {author} {\bibinfo {author} {\bibfnamefont {S.}~\bibnamefont
  {Middleman}},\ }\bibfield  {title} {\emph {\enquote {\bibinfo {title} {Effect
  of molecular weight distribution on viscosity of polymeric fluids},}\ }}\href
  {\doibase 10.1002/app.1967.070110309} {\bibfield  {journal} {\bibinfo
  {journal} {J. Appl. Polym. Sci.}\ }\textbf {\bibinfo {volume} {11}},\
  \bibinfo {pages} {417} (\bibinfo {year} {1967})}\BibitemShut {NoStop}%
\balance
\bibitem [{\citenamefont {Bersted}(1979)}]{Bersted1979}%
  \BibitemOpen
  \bibfield  {author} {\bibinfo {author} {\bibfnamefont {B.~H.}\ \bibnamefont
  {Bersted}},\ }\bibfield  {title} {\emph {\enquote {\bibinfo {title} {Effect
  of molecular weight distribution on elongational viscosity of undiluted
  polymer fluids},}\ }}\href {\doibase 10.1002/app.1979.070240305} {\bibfield
  {journal} {\bibinfo  {journal} {J. Appl. Polym. Sci.}\ }\textbf {\bibinfo
  {volume} {24}},\ \bibinfo {pages} {671} (\bibinfo {year} {1979})}\BibitemShut
  {NoStop}%
\bibitem [{\citenamefont {Gentekos}\ \emph {et~al.}(2019)\citenamefont
  {Gentekos}, \citenamefont {Sifri},\ and\ \citenamefont
  {Fors}}]{Gentekos2019}%
  \BibitemOpen
  \bibfield  {author} {\bibinfo {author} {\bibfnamefont {D.~T.}\ \bibnamefont
  {Gentekos}}, \bibinfo {author} {\bibfnamefont {R.~J.}\ \bibnamefont {Sifri}},
  \ and\ \bibinfo {author} {\bibfnamefont {B.~P.}\ \bibnamefont {Fors}},\
  }\bibfield  {title} {\emph {\enquote {\bibinfo {title} {Controlling polymer
  properties through the shape of the molecular-weight distribution},}\ }}\href
  {\doibase 10.1038/s41578-019-0138-8} {\bibfield  {journal} {\bibinfo
  {journal} {Nat. Rev. Mater.}\ }\textbf {\bibinfo {volume} {4}},\ \bibinfo
  {pages} {761} (\bibinfo {year} {2019})}\BibitemShut {NoStop}%
\bibitem [{\citenamefont {Plog}\ \emph {et~al.}(2005)\citenamefont {Plog},
  \citenamefont {Kulicke},\ and\ \citenamefont {Clasen}}]{Plog2005}%
  \BibitemOpen
  \bibfield  {author} {\bibinfo {author} {\bibfnamefont {J.~P.}\ \bibnamefont
  {Plog}}, \bibinfo {author} {\bibfnamefont {W.-M.}\ \bibnamefont {Kulicke}}, \
  and\ \bibinfo {author} {\bibfnamefont {C.}~\bibnamefont {Clasen}},\
  }\bibfield  {title} {\emph {\enquote {\bibinfo {title} {Influence of the
  molar mass distribution on the elongational behaviour of polymer solutions in
  capillary breakup},}\ }}\href {\doibase 10.1515/arh-2005-0002} {\bibfield
  {journal} {\bibinfo  {journal} {Appl. Rheol.}\ }\textbf {\bibinfo {volume}
  {15}},\ \bibinfo {pages} {28} (\bibinfo {year} {2005})}\BibitemShut {NoStop}%
\bibitem [{\citenamefont {Kim}\ and\ \citenamefont {Lee}(2019)}]{Kim2019}%
  \BibitemOpen
  \bibfield  {author} {\bibinfo {author} {\bibfnamefont {S.~G.}\ \bibnamefont
  {Kim}}\ and\ \bibinfo {author} {\bibfnamefont {H.~S.}\ \bibnamefont {Lee}},\
  }\bibfield  {title} {\emph {\enquote {\bibinfo {title} {Concentration
  dependence of the extensional relaxation time and finite extensibility in
  dilute and semidilute polymer solutions using a microfluidic rheometer},}\
  }}\href {\doibase 10.1021/acs.macromol.9b01143} {\bibfield  {journal}
  {\bibinfo  {journal} {Macromolecules}\ }\textbf {\bibinfo {volume} {52}},\
  \bibinfo {pages} {9585} (\bibinfo {year} {2019})}\BibitemShut {NoStop}%
\bibitem [{\citenamefont {Palangetic}\ \emph {et~al.}(2014)\citenamefont
  {Palangetic}, \citenamefont {Reddy}, \citenamefont {Srinivasan},
  \citenamefont {Cohen}, \citenamefont {McKinley},\ and\ \citenamefont
  {Clasen}}]{Palangetic2014}%
  \BibitemOpen
  \bibfield  {author} {\bibinfo {author} {\bibfnamefont {L.}~\bibnamefont
  {Palangetic}}, \bibinfo {author} {\bibfnamefont {N.~K.}\ \bibnamefont
  {Reddy}}, \bibinfo {author} {\bibfnamefont {S.}~\bibnamefont {Srinivasan}},
  \bibinfo {author} {\bibfnamefont {R.~E.}\ \bibnamefont {Cohen}}, \bibinfo
  {author} {\bibfnamefont {G.~H.}\ \bibnamefont {McKinley}}, \ and\ \bibinfo
  {author} {\bibfnamefont {C.}~\bibnamefont {Clasen}},\ }\bibfield  {title}
  {\emph {\enquote {\bibinfo {title} {Dispersity and spinnability: {W}hy highly
  polydisperse polymer solutions are desirable for electrospinning},}\ }}\href
  {\doibase 10.1016/j.polymer.2014.07.047} {\bibfield  {journal} {\bibinfo
  {journal} {Polymer}\ }\textbf {\bibinfo {volume} {55}},\ \bibinfo {pages}
  {4920} (\bibinfo {year} {2014})}\BibitemShut {NoStop}%
\bibitem [{\citenamefont {Merchiers}\ \emph {et~al.}(2022)\citenamefont
  {Merchiers}, \citenamefont {Reddy},\ and\ \citenamefont
  {Sharma}}]{Merchiers2022}%
  \BibitemOpen
  \bibfield  {author} {\bibinfo {author} {\bibfnamefont {J.}~\bibnamefont
  {Merchiers}}, \bibinfo {author} {\bibfnamefont {N.~K.}\ \bibnamefont
  {Reddy}}, \ and\ \bibinfo {author} {\bibfnamefont {V.}~\bibnamefont
  {Sharma}},\ }\bibfield  {title} {\emph {\enquote {\bibinfo {title}
  {Extensibility-enriched spinnability and enhanced sorption and strength of
  centrifugally spun polystyrene fiber mats},}\ }}\href {\doibase
  10.1021/acs.macromol.1c02164} {\bibfield  {journal} {\bibinfo  {journal}
  {Macromolecules}\ }\textbf {\bibinfo {volume} {55}},\ \bibinfo {pages} {942}
  (\bibinfo {year} {2022})}\BibitemShut {NoStop}%
\bibitem [{\citenamefont {Haward}\ \emph {et~al.}(2016)\citenamefont {Haward},
  \citenamefont {McKinley},\ and\ \citenamefont {Shen}}]{Haward2016c}%
  \BibitemOpen
  \bibfield  {author} {\bibinfo {author} {\bibfnamefont {S.~J.}\ \bibnamefont
  {Haward}}, \bibinfo {author} {\bibfnamefont {G.~H.}\ \bibnamefont
  {McKinley}}, \ and\ \bibinfo {author} {\bibfnamefont {A.~Q.}\ \bibnamefont
  {Shen}},\ }\bibfield  {title} {\emph {\enquote {\bibinfo {title} {Elastic
  instabilities in planar elongational flow of monodisperse polymer
  solutions},}\ }}\href {\doibase 10.1038/srep33029} {\bibfield  {journal}
  {\bibinfo  {journal} {Scientific Reports}\ }\textbf {\bibinfo {volume} {6}},\
  \bibinfo {pages} {33029} (\bibinfo {year} {2016})}\BibitemShut {NoStop}%
\bibitem [{\citenamefont {Berry}(1967)}]{Berry1967}%
  \BibitemOpen
  \bibfield  {author} {\bibinfo {author} {\bibfnamefont {G.~C.}\ \bibnamefont
  {Berry}},\ }\bibfield  {title} {\emph {\enquote {\bibinfo {title}
  {Thermodynamic and conformational properties of polystyrene. {II}.
  {I}ntrinsic viscosity studies on dilute solutions of linear polystyrene},}\
  }}\href {\doibase 10.1063/1.1840854} {\bibfield  {journal} {\bibinfo
  {journal} {J. Chem. Phys.}\ }\textbf {\bibinfo {volume} {46}},\ \bibinfo
  {pages} {1338} (\bibinfo {year} {1967})}\BibitemShut {NoStop}%
\bibitem [{\citenamefont {Brandrup}\ \emph {et~al.}(2003)\citenamefont
  {Brandrup}, \citenamefont {Immergut},\ and\ \citenamefont
  {Grulke}}]{Polymer_Handbook}%
  \BibitemOpen
  \bibinfo {editor} {\bibfnamefont {J.}~\bibnamefont {Brandrup}}, \bibinfo
  {editor} {\bibfnamefont {E.~H.}\ \bibnamefont {Immergut}}, \ and\ \bibinfo
  {editor} {\bibfnamefont {E.~A.}\ \bibnamefont {Grulke}},\ eds.,\ \href@noop
  {} {\emph {\bibinfo {title} {Polymer Handbook}}},\ \bibinfo {edition} {4th}\
  ed.\ (\bibinfo  {publisher} {Wiley Interscience},\ \bibinfo {address}
  {Hoboken, New Jersey, USA},\ \bibinfo {year} {2003})\BibitemShut {NoStop}%
\bibitem [{\citenamefont {Graessley}(1980)}]{Graessley1980}%
  \BibitemOpen
  \bibfield  {author} {\bibinfo {author} {\bibfnamefont {W.~W.}\ \bibnamefont
  {Graessley}},\ }\bibfield  {title} {\emph {\enquote {\bibinfo {title}
  {Polymer chain dimensions and the dependence of viscoelastic properties on
  concentration, molecular weight and solvent power},}\ }}\href {\doibase
  10.1016/0032-3861(80)90266-9} {\bibfield  {journal} {\bibinfo  {journal}
  {Polymer}\ }\textbf {\bibinfo {volume} {21}},\ \bibinfo {pages} {258}
  (\bibinfo {year} {1980})}\BibitemShut {NoStop}%
\bibitem [{\citenamefont {Zimm}(1956)}]{Zimm1956}%
  \BibitemOpen
  \bibfield  {author} {\bibinfo {author} {\bibfnamefont {B.~H.}\ \bibnamefont
  {Zimm}},\ }\bibfield  {title} {\emph {\enquote {\bibinfo {title} {Dynamics of
  polymer molecules in dilute solution: {V}iscoelasticity, flow birefringence
  and dielectric loss},}\ }}\href {\doibase 10.1063/1.1742462} {\bibfield
  {journal} {\bibinfo  {journal} {J. Chem. Phys.}\ }\textbf {\bibinfo {volume}
  {24}},\ \bibinfo {pages} {269} (\bibinfo {year} {1956})}\BibitemShut
  {NoStop}%
\bibitem [{\citenamefont {Rubinstein}\ and\ \citenamefont
  {Colby}(2003)}]{Rubinstein_Colby}%
  \BibitemOpen
  \bibfield  {author} {\bibinfo {author} {\bibfnamefont {M.}~\bibnamefont
  {Rubinstein}}\ and\ \bibinfo {author} {\bibfnamefont {R.~H.}\ \bibnamefont
  {Colby}},\ }\href@noop {} {\emph {\bibinfo {title} {Polymer Physics}}}\
  (\bibinfo  {publisher} {Oxford University Press},\ \bibinfo {address} {New
  York, USA},\ \bibinfo {year} {2003})\BibitemShut {NoStop}%
\bibitem [{\citenamefont {Ricci}\ \emph {et~al.}(1986)\citenamefont {Ricci},
  \citenamefont {Sangiorgi},\ and\ \citenamefont {Passerone}}]{Ricci1986}%
  \BibitemOpen
  \bibfield  {author} {\bibinfo {author} {\bibfnamefont {E.}~\bibnamefont
  {Ricci}}, \bibinfo {author} {\bibfnamefont {R.}~\bibnamefont {Sangiorgi}}, \
  and\ \bibinfo {author} {\bibfnamefont {A.}~\bibnamefont {Passerone}},\
  }\bibfield  {title} {\emph {\enquote {\bibinfo {title} {Density and surface
  tension of dioctylphthalate, silicone oil and their solutions},}\ }}\href
  {\doibase 10.1016/0257-8972(86)90060-5} {\bibfield  {journal} {\bibinfo
  {journal} {Surf. Coat. Tech.}\ }\textbf {\bibinfo {volume} {28}},\ \bibinfo
  {pages} {215} (\bibinfo {year} {1986})}\BibitemShut {NoStop}%
\bibitem [{\citenamefont {Doi}\ and\ \citenamefont {Edwards}(1986)}]{Doi1986}%
  \BibitemOpen
  \bibfield  {author} {\bibinfo {author} {\bibfnamefont {M.}~\bibnamefont
  {Doi}}\ and\ \bibinfo {author} {\bibfnamefont {S.}~\bibnamefont {Edwards}},\
  }\href@noop {} {\emph {\bibinfo {title} {The Theory of Polymer Dynamics}}}\
  (\bibinfo  {publisher} {Oxford University Press},\ \bibinfo {address} {New
  York},\ \bibinfo {year} {1986})\BibitemShut {NoStop}%
\bibitem [{\citenamefont {Larson}(2005)}]{Larson2005}%
  \BibitemOpen
  \bibfield  {author} {\bibinfo {author} {\bibfnamefont {R.~G.}\ \bibnamefont
  {Larson}},\ }\bibfield  {title} {\emph {\enquote {\bibinfo {title} {The
  rheology of dilute solutions of flexible polymers: {P}rogress and
  problems},}\ }}\href {\doibase 10.1122/1.1835336} {\bibfield  {journal}
  {\bibinfo  {journal} {J. Rheol.}\ }\textbf {\bibinfo {volume} {49}},\
  \bibinfo {pages} {1} (\bibinfo {year} {2005})}\BibitemShut {NoStop}%
\bibitem [{\citenamefont {de~Gennes}(1974)}]{DeGennes1974}%
  \BibitemOpen
  \bibfield  {author} {\bibinfo {author} {\bibfnamefont {P.~G.}\ \bibnamefont
  {de~Gennes}},\ }\bibfield  {title} {\emph {\enquote {\bibinfo {title}
  {Coil-stretch transition of dilute flexible polymers under ultrahigh velocity
  gradients},}\ }}\href {\doibase 10.1063/1.1681018} {\bibfield  {journal}
  {\bibinfo  {journal} {J. Chem. Phys.}\ }\textbf {\bibinfo {volume} {60}},\
  \bibinfo {pages} {5030} (\bibinfo {year} {1974})}\BibitemShut {NoStop}%
\bibitem [{\citenamefont {Hinch}(1977)}]{Hinch1977}%
  \BibitemOpen
  \bibfield  {author} {\bibinfo {author} {\bibfnamefont {E.~J.}\ \bibnamefont
  {Hinch}},\ }\bibfield  {title} {\emph {\enquote {\bibinfo {title} {Mechanical
  models of dilute polymer solutions in strong flows},}\ }}\href {\doibase
  10.1063/1.861735} {\bibfield  {journal} {\bibinfo  {journal} {Phys. Fluids}\
  }\textbf {\bibinfo {volume} {20}},\ \bibinfo {pages} {S22} (\bibinfo {year}
  {1977})}\BibitemShut {NoStop}%
\bibitem [{\citenamefont {McKinley}\ and\ \citenamefont
  {Sridhar}(2002)}]{McKinley2002}%
  \BibitemOpen
  \bibfield  {author} {\bibinfo {author} {\bibfnamefont {G.~H.}\ \bibnamefont
  {McKinley}}\ and\ \bibinfo {author} {\bibfnamefont {T.}~\bibnamefont
  {Sridhar}},\ }\bibfield  {title} {\emph {\enquote {\bibinfo {title} {Filament
  stretching rheometry of complex fluids},}\ }}\href@noop {} {\bibfield
  {journal} {\bibinfo  {journal} {Annu. Rev. Fluid Mech.}\ }\textbf {\bibinfo
  {volume} {34}},\ \bibinfo {pages} {375} (\bibinfo {year} {2002})}\BibitemShut
  {NoStop}%
\bibitem [{\citenamefont {Bazilevsky}\ \emph {et~al.}(1990)\citenamefont
  {Bazilevsky}, \citenamefont {Entov},\ and\ \citenamefont
  {Rozhkov}}]{Bazilevsky1990}%
  \BibitemOpen
  \bibfield  {author} {\bibinfo {author} {\bibfnamefont {A.}~\bibnamefont
  {Bazilevsky}}, \bibinfo {author} {\bibfnamefont {V.}~\bibnamefont {Entov}}, \
  and\ \bibinfo {author} {\bibfnamefont {A.}~\bibnamefont {Rozhkov}},\ }in\
  \href@noop {} {\emph {\bibinfo {booktitle} {Proceedings of the Third European
  Rheology Conference}}},\ \bibinfo {editor} {edited by\ \bibinfo {editor}
  {\bibfnamefont {D.~R.}\ \bibnamefont {Oliver}}}\ (\bibinfo  {publisher}
  {Elsevier},\ \bibinfo {address} {New York},\ \bibinfo {year} {1990})\ pp.\
  \bibinfo {pages} {41--43}\BibitemShut {NoStop}%
\bibitem [{\citenamefont {Gaillard}\ \emph {et~al.}(2024)\citenamefont
  {Gaillard}, \citenamefont {Herrada}, \citenamefont {Deblais}, \citenamefont
  {van Poelgeest}, \citenamefont {Laruelle}, \citenamefont {Eggers},\ and\
  \citenamefont {Bonn}}]{Gaillard2024}%
  \BibitemOpen
  \bibfield  {author} {\bibinfo {author} {\bibfnamefont {A.}~\bibnamefont
  {Gaillard}}, \bibinfo {author} {\bibfnamefont {M.~A.}\ \bibnamefont
  {Herrada}}, \bibinfo {author} {\bibfnamefont {A.}~\bibnamefont {Deblais}},
  \bibinfo {author} {\bibfnamefont {C.}~\bibnamefont {van Poelgeest}}, \bibinfo
  {author} {\bibfnamefont {L.}~\bibnamefont {Laruelle}}, \bibinfo {author}
  {\bibfnamefont {J.}~\bibnamefont {Eggers}}, \ and\ \bibinfo {author}
  {\bibfnamefont {D.}~\bibnamefont {Bonn}},\ }\href@noop {} {\enquote {\bibinfo
  {title} {When does the elastic regime begin in viscoelastic pinch-off?}}\ }
  (\bibinfo {year} {2024}),\ \Eprint {http://arxiv.org/abs/2406.02303v1}
  {arXiv:2406.02303v1 [cond-mat.soft]} \BibitemShut {NoStop}%
\bibitem [{\citenamefont {Keller}\ and\ \citenamefont
  {Odell}(1985)}]{Keller1985}%
  \BibitemOpen
  \bibfield  {author} {\bibinfo {author} {\bibfnamefont {A.}~\bibnamefont
  {Keller}}\ and\ \bibinfo {author} {\bibfnamefont {J.~A.}\ \bibnamefont
  {Odell}},\ }\bibfield  {title} {\emph {\enquote {\bibinfo {title} {The
  extensibility of macromolecules in solution; {A} new focus for macromolecular
  science},}\ }}\href {\doibase 10.1007/BF01415506} {\bibfield  {journal}
  {\bibinfo  {journal} {Colloid Polym. Sci.}\ }\textbf {\bibinfo {volume}
  {263}},\ \bibinfo {pages} {181} (\bibinfo {year} {1985})}\BibitemShut
  {NoStop}%
\bibitem [{\citenamefont {Rabin}(1985)}]{Rabin1985}%
  \BibitemOpen
  \bibfield  {author} {\bibinfo {author} {\bibfnamefont {Y.}~\bibnamefont
  {Rabin}},\ }\bibfield  {title} {\emph {\enquote {\bibinfo {title} {On the
  universality of ideal {Z}imm dynamics of polymers in extensional flows},}\
  }}\href {\doibase 10.1002/pol.1985.130230103} {\bibfield  {journal} {\bibinfo
   {journal} {J. Polym. Sci. Polym. Lett. Ed.}\ }\textbf {\bibinfo {volume}
  {23}},\ \bibinfo {pages} {11} (\bibinfo {year} {1985})}\BibitemShut {NoStop}%
\bibitem [{\citenamefont {Perkins}\ \emph {et~al.}(1997)\citenamefont
  {Perkins}, \citenamefont {Smith},\ and\ \citenamefont {Chu}}]{Perkins1997}%
  \BibitemOpen
  \bibfield  {author} {\bibinfo {author} {\bibfnamefont {T.~T.}\ \bibnamefont
  {Perkins}}, \bibinfo {author} {\bibfnamefont {D.~E.}\ \bibnamefont {Smith}},
  \ and\ \bibinfo {author} {\bibfnamefont {S.}~\bibnamefont {Chu}},\ }\bibfield
   {title} {\emph {\enquote {\bibinfo {title} {Single polymer dynamics in an
  elongational flow},}\ }}\href {\doibase 10.1126/science.276.5321.2016}
  {\bibfield  {journal} {\bibinfo  {journal} {Science}\ }\textbf {\bibinfo
  {volume} {276}},\ \bibinfo {pages} {2016} (\bibinfo {year}
  {1997})}\BibitemShut {NoStop}%
\bibitem [{\citenamefont {Ferry}(1980)}]{Ferry1980}%
  \BibitemOpen
  \bibfield  {author} {\bibinfo {author} {\bibfnamefont {J.~D.}\ \bibnamefont
  {Ferry}},\ }\href@noop {} {\emph {\bibinfo {title} {Viscoelastic Properties
  of Polymers}}},\ \bibinfo {edition} {3rd}\ ed.\ (\bibinfo  {publisher} {John
  Wiley \& Sons, Inc.},\ \bibinfo {address} {New York},\ \bibinfo {year}
  {1980})\BibitemShut {NoStop}%
\bibitem [{\citenamefont {Farrell}\ \emph {et~al.}(1980)\citenamefont
  {Farrell}, \citenamefont {Keller}, \citenamefont {Miles},\ and\ \citenamefont
  {Pope}}]{Farrell1980}%
  \BibitemOpen
  \bibfield  {author} {\bibinfo {author} {\bibfnamefont {C.~J.}\ \bibnamefont
  {Farrell}}, \bibinfo {author} {\bibfnamefont {A.}~\bibnamefont {Keller}},
  \bibinfo {author} {\bibfnamefont {M.~J.}\ \bibnamefont {Miles}}, \ and\
  \bibinfo {author} {\bibfnamefont {D.~P.}\ \bibnamefont {Pope}},\ }\bibfield
  {title} {\emph {\enquote {\bibinfo {title} {Conformational relaxation time in
  polymer solutions by elongational flow experiments: 1. {D}etermination of
  extensional relaxation time and its molecular weight dependence},}\ }}\href
  {\doibase 10.1016/0032-3861(80)90195-0} {\bibfield  {journal} {\bibinfo
  {journal} {Polymer}\ }\textbf {\bibinfo {volume} {21}},\ \bibinfo {pages}
  {1292} (\bibinfo {year} {1980})}\BibitemShut {NoStop}%
\bibitem [{\citenamefont {Fuller}\ and\ \citenamefont
  {Leal}(1980)}]{Fuller1980}%
  \BibitemOpen
  \bibfield  {author} {\bibinfo {author} {\bibfnamefont {G.~G.}\ \bibnamefont
  {Fuller}}\ and\ \bibinfo {author} {\bibfnamefont {L.~G.}\ \bibnamefont
  {Leal}},\ }\bibfield  {title} {\emph {\enquote {\bibinfo {title} {Flow
  birefringence of dilute polymer solutions in two-dimensional flows},}\
  }}\href {\doibase 10.1007/BF01517512} {\bibfield  {journal} {\bibinfo
  {journal} {Rheol. Acta}\ }\textbf {\bibinfo {volume} {19}},\ \bibinfo {pages}
  {580} (\bibinfo {year} {1980})}\BibitemShut {NoStop}%
\bibitem [{\citenamefont {Rabin}\ \emph {et~al.}(1985)\citenamefont {Rabin},
  \citenamefont {Henyey},\ and\ \citenamefont {Pathria}}]{Rabin1985b}%
  \BibitemOpen
  \bibfield  {author} {\bibinfo {author} {\bibfnamefont {Y.}~\bibnamefont
  {Rabin}}, \bibinfo {author} {\bibfnamefont {F.~S.}\ \bibnamefont {Henyey}}, \
  and\ \bibinfo {author} {\bibfnamefont {R.~K.}\ \bibnamefont {Pathria}},\
  }\bibfield  {title} {\emph {\enquote {\bibinfo {title} {Scaling behavior of
  dilute polymer solutions in elongational flows},}\ }}\href {\doibase
  10.1103/PhysRevLett.55.201} {\bibfield  {journal} {\bibinfo  {journal} {Phys.
  Rev. Lett.}\ }\textbf {\bibinfo {volume} {55}},\ \bibinfo {pages} {201}
  (\bibinfo {year} {1985})}\BibitemShut {NoStop}%
\bibitem [{\citenamefont {Calabrese}\ \emph {et~al.}(2024)\citenamefont
  {Calabrese}, \citenamefont {Shen},\ and\ \citenamefont
  {Haward}}]{Calabrese2024}%
  \BibitemOpen
  \bibfield  {author} {\bibinfo {author} {\bibfnamefont {V.}~\bibnamefont
  {Calabrese}}, \bibinfo {author} {\bibfnamefont {A.~Q.}\ \bibnamefont {Shen}},
  \ and\ \bibinfo {author} {\bibfnamefont {S.~J.}\ \bibnamefont {Haward}},\
  }\href@noop {} {\enquote {\bibinfo {title} {How do polymers stretch in
  capillary-driven extensional flows?}}\ } (\bibinfo {year} {2024}),\ \Eprint
  {http://arxiv.org/abs/2403.04103v1} {arXiv:2403.04103v1 [cond-mat.soft]}
  \BibitemShut {NoStop}%
\end{thebibliography}

%

\newpage

\section*{\label{ESI}SUPPLEMENTARY INFORMATION}
\vspace{-0.1in}
\renewcommand{\thefigure}{S\arabic{figure}}
\setcounter{figure}{0}

\renewcommand{\thesubsection}{S\arabic{subsection}}

\subsection{Steady flow curves}
\begin{figure}[H]
    \centering
    \includegraphics[width=6.5cm]{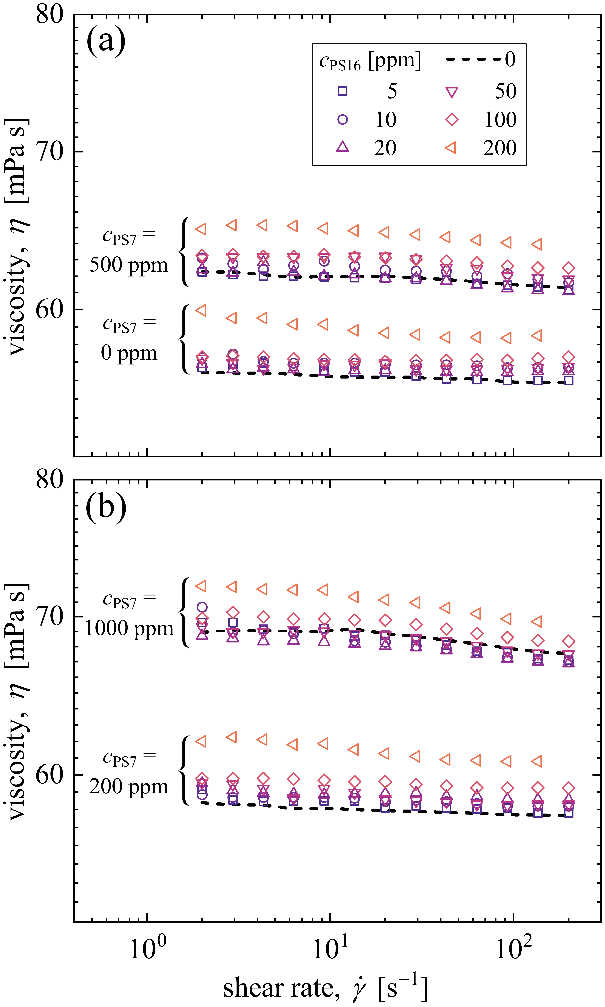}
    \caption{Steady shear flow curves  of all the tested polymer solutions. Part (a) shows data for PS16 at various concentrations dissolved in pure solvent (i.e., $c_{\text{PS7}} = 0~\text{ppm}$) and mixed with PS7 at 500~ppm. Part (b) shows data for PS16 at various concentrations mixed with PS7 at 200~ppm and 1000~ppm.
 }
    \label{flowcurves}
\end{figure}

\vspace{-0.1in}
The polymeric test solutions used in the study (i.e., pure PS7 and PS16 solutions and their blends) are characterized in steady shear at 25$^{\circ}$C using an Anton-Paar MCR502 stress-controlled rotational rheometer fitted with a 50~mm diameter 1$^{\circ}$ cone-and-plate geometry (see Fig.~\ref{flowcurves}). As expected for dilute polymer solutions, the fluids have viscosities $\eta$ close to that of the solvent $\eta_s \approx 57~\text{mPa~s}$ (the minimum solvent-to-total viscosity ratio is $\eta_s / \eta \approx 0.8$), and none of the fluids exhibit more than a weak degree of shear thinning. 

\subsection{Effect of plate size}
\begin{figure}[H]
    \centering
    \includegraphics[width=6.3cm]{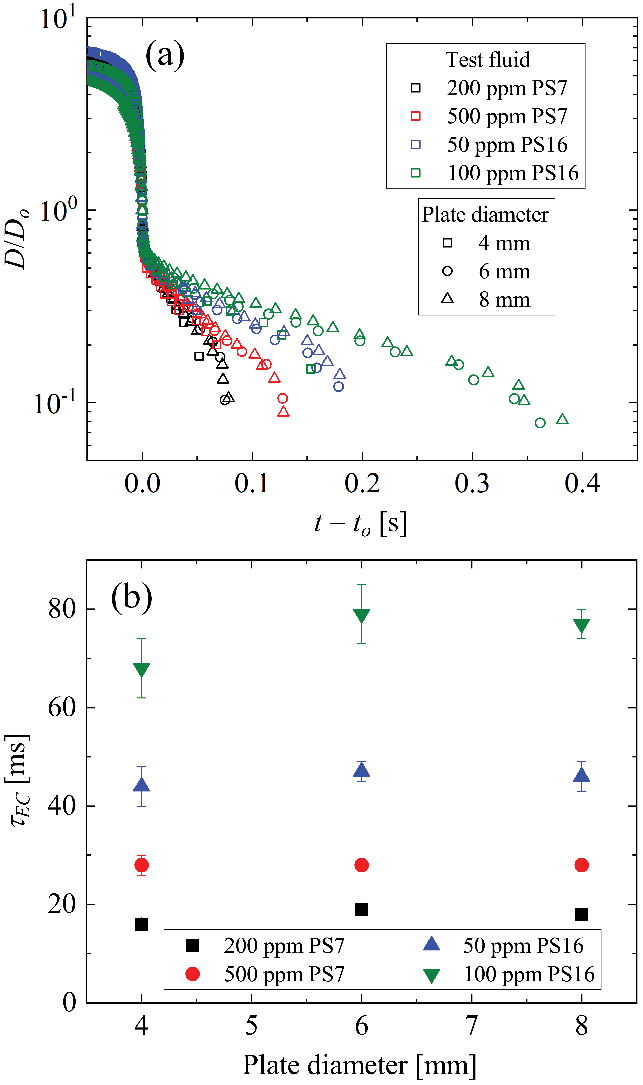}
\vspace{-0.1in}
    \caption{Effect of plate size on the capillary thinning of dilute polymer solutions. (a) Representative normalized diameter \emph{versus} time profiles for various polymer solutions using 4~mm, 6~mm and 8~mm diameter end plates fitted on the CaBER instrument ($D_o$ is the diameter at time $t=t_o$). (b) timescale $\tau_{EC}$ in the exponential elastocapillary thinning regime as a function of the plate size. Error bars represent the standard deviation over five repeated measurements.
 }
    \label{platesize}
\end{figure}

\vspace{-0.1in}
Due to recent reports that the measurement of $\tau_{EC}$ by the SRM method can depend on the plate diameter \cite{Gaillard2023,Gaillard2024}, we tested a number of our fluids with plates of 4~mm and 8~mm diameter, as well as with the 6~mm diameter plates we used as standard in our experiments presented in the main text. In contrary to those recent reports, we observed no systematic variation of the EC thinning dynamics using these different plates (see Fig.~\ref{platesize}).

\vspace{-0.3in}

\subsection{Strain rate at time $t_o$}
\begin{figure}[H]
    \centering
\vspace{-0.1in}
    \includegraphics[width=6.3cm]{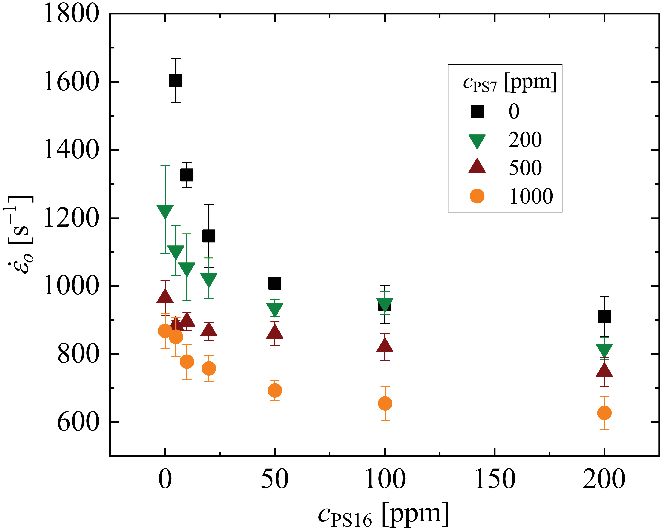}
\vspace{-0.1in}
    \caption{Instantaneous strain rate $\dot\varepsilon_o$ at the onset of the EC regime at time $t=t_o$. Data is presented for all of the examined test solutions as a function of the PS16 concentration.
 }
    \label{eps-dot_o}
\end{figure}

\vspace{-0.1in}
In Fig.~\ref{eps-dot_o} we present the peak extensional strain rate $\dot\varepsilon_o$ measured at time $t=t_o$ (i.e., at the onset of the EC thinning regime), showing a general reduction with increasing polymer concentration as expected (see Sec.~III.D. of the main text).

\end{document}